\documentclass[epj]{svjour}
\usepackage{graphics}
\begin{document}
\title{Measurement of the reaction ${\bf \gamma p \rightarrow K^0 \Sigma^+}$ 
at photon energies up to ${\bf 2.6}$ GeV\thanks{This work is supported in part 
by the Deutsche Forschungsgemeinschaft (DFG) (SPP KL 980/2-3)}}
\titlerunning{Measurement of the reaction $\gamma p \rightarrow K^0 \Sigma^+$ 
at photon energies up to $2.6$ GeV}
\author{
R. Lawall\inst{1,}\thanks{Part of doctoral thesis (R.Lawall, doctoral thesis,
\newline Bonn University (2004), Bonn-IR-2004-01), http://saphir.
\newline physik.uni-bonn.de/saphir/thesis.html}, 
J. Barth\inst{1},
C. Bennhold\inst{3},
K.-H. Glander\inst{1}, 
S. Goers\inst{1},
J. Hannappel\inst{1},
N. J\"open\inst{1}, 
F. Klein\inst{1,}\thanks{email: klein@physik.uni-bonn.de},
E. Klempt\inst{2}, 
T. Mart\inst{4},
D. Menze\inst{1}, 
M. Ostrick\inst{1}, 
E. Paul\inst{1}, 
I. Schulday\inst{1}, 
W.J. Schwille\inst{1}, 
F.W. Wieland\inst{1}, 
C. Wu\inst{1}
% \thanks is optional - remove next line if not needed
%\thanks{\emph{Present address:} Insert the address here if needed}%
}                     % Do not remove
\authorrunning{R.\,Lawall et al.}
%
%\offprints{Offprint}          % Insert a name or remove this line
%
%\institute{Insert the first address here \and the second here}
\institute{
Physikalisches Institut der Universit\"at Bonn, Germany 
\and Helmholtz-Institut f\"ur Strahlen- und Kernphysik Bonn, Germany
\and Dep. of Physics, George Washington University, Washington DC, USA
\and Departemen Fisika, FMIPA, Universitas Indonesia, Depok 16424, Indonesia
%\and No longer working at this experiment
}
\date{Received: date / Revised version: date}
% The correct dates will be entered by Springer
%
\abstract{
%Insert your abstract here.
The reaction $\gamma p \rightarrow K^0 \Sigma^+$ was measured in 
the photon energy range from threshold up to 2.6 GeV with the 
SAPHIR detector at the electron stretcher facility, ELSA, 
in Bonn. Results are presented on the reaction cross section 
and the polarization of the $\Sigma^+$ as a function of the 
kaon production angle in the centre-of-mass system, $\cos\Theta_K^{c.m.}$, 
and the photon energy. The cross section is lower and varies less with photon energy 
and kaon production angle than that of $\gamma p \rightarrow K^+\Sigma^0$. 
The $\Sigma^+$ is polarized predominantly at $\cos\Theta_K^{c.m.} \approx 0$.
The data presented here are 
more precise than previous ones obtained with SAPHIR and 
extend the photon energy range to higher values. They are compared to 
isobar model calculations.
\PACS{
      {PACS-key}{describing text of that key}   \and
      {PACS-key}{describing text of that key}
     } % end of PACS codes
} %end of abstract
\maketitle
\parindent0mm
\begin{sloppypar}

\section{Introduction}
\label{intro}

Hadronic final states produced in hadron- or photon-indu-\\
ced reactions on nucleons at low energies are investigated with 
respect to production mechanisms, in particular to the formation of 
baryonic resonances which decay into hadrons. Most of the known nucleon 
and delta resonances were discovered in this way. However, model 
calculations (\cite{theory1}, \cite{theory2}) predict more resonant 
states than have been observed in experiments so far. Some of the missing 
non-strange resonances are predicted to decay into final states comprising 
strange particle pairs.\\
The SAPHIR experiment has measured such final states. Results on the reactions 
$\gamma p \rightarrow K^+ \Lambda$ and $\gamma p \rightarrow K^+ \Sigma^0$ were 
published elsewhere \cite{Glan03b}. Here we report about the reaction $
\gamma p \rightarrow K^0 \Sigma^+$ which adds another 
isospin configuration of the $K \Sigma$ final state. \\
Results on $\gamma p \rightarrow K^0 \Sigma^+$ based on 30 million triggers 
from earlier data taking with SAPHIR were published \cite{Goers99}. 
The new results presented here are based on 180 million triggers which were taken 
with an upgraded SAPHIR detector setup in an extended photon energy range from 
reaction threshold up to 2.6 GeV. This measurement of cross sections is more reliable 
than the previous one, since it has been found that, in the previous analysis, background 
from other reactions was underestimated and not sufficiently eliminated from the event 
sample of the reaction $\gamma p \rightarrow K^0 \Sigma^+$. In the current analysis this 
background has been removed thoroughly. The new cross sections show systematically lower values.\\
The data are available via internet.\footnote{http://saphir.physik.uni-bonn.de/saphir/publications}

\section{The experiment}
\label{saphir}

\begin{figure*}[ht!]
\hspace{4cm}
\resizebox{0.6\textwidth}{!}{%
  \rotatebox{270}{\includegraphics{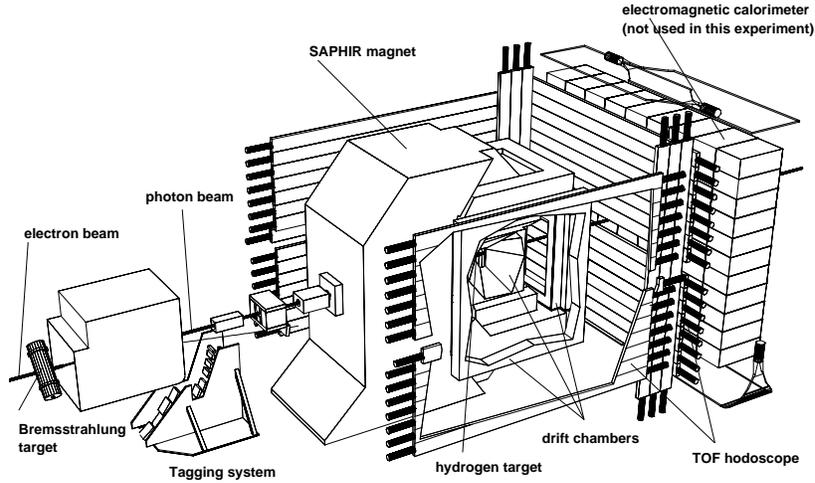}}
}
\caption{Sketch of the SAPHIR detector.}
\label{pic:saphir}       % Give a unique label
\end{figure*}
Data were taken with the magnetic multiparticle spectrometer SAPHIR \cite{Schwille} at 
the 3.5 GeV electron stretcher facility ELSA \cite{Husmann}. The setup is shown 
schematically in fig. \ref{pic:saphir}. An extracted electron beam of 2.8 GeV was directed 
on a radiator target to provide an energy-tagged photon beam within the range from 
0.868 to 2.650 GeV. The data taking was based on a trigger defined by a coincidence of 
signals from the scattered electrons in the tagging system with at least two charged 
particles in the scintillator hodoscopes and no signal from a beam-veto-counter downstream 
of SAPHIR which detected non-interacting photons.\\
The drift chambers served to measure charged particles. The scintillator hodoscopes 
were used for time-of-flight (TOF) measurements which allowed together with the track 
momentum (measured with the drift chambers in the magnetic field) to calculate 
the particle masses.\\
The experimental set-up of SAPHIR was upgraded for this data taking. A new tagging system 
\cite{tof} with improved energy resolution and extended electron energy range was installed. 
A planar drift chamber was added to the central drift chamber which improved the track 
reconstruction in forward direction \cite{Glan03a}. The data analyzed here stem from four 
data taking periods in 1997 and 1998.

\section{Event reconstruction and event selection}
\label{selection}
\begin{figure}[b!]
\resizebox{0.5\textwidth}{!}{\includegraphics{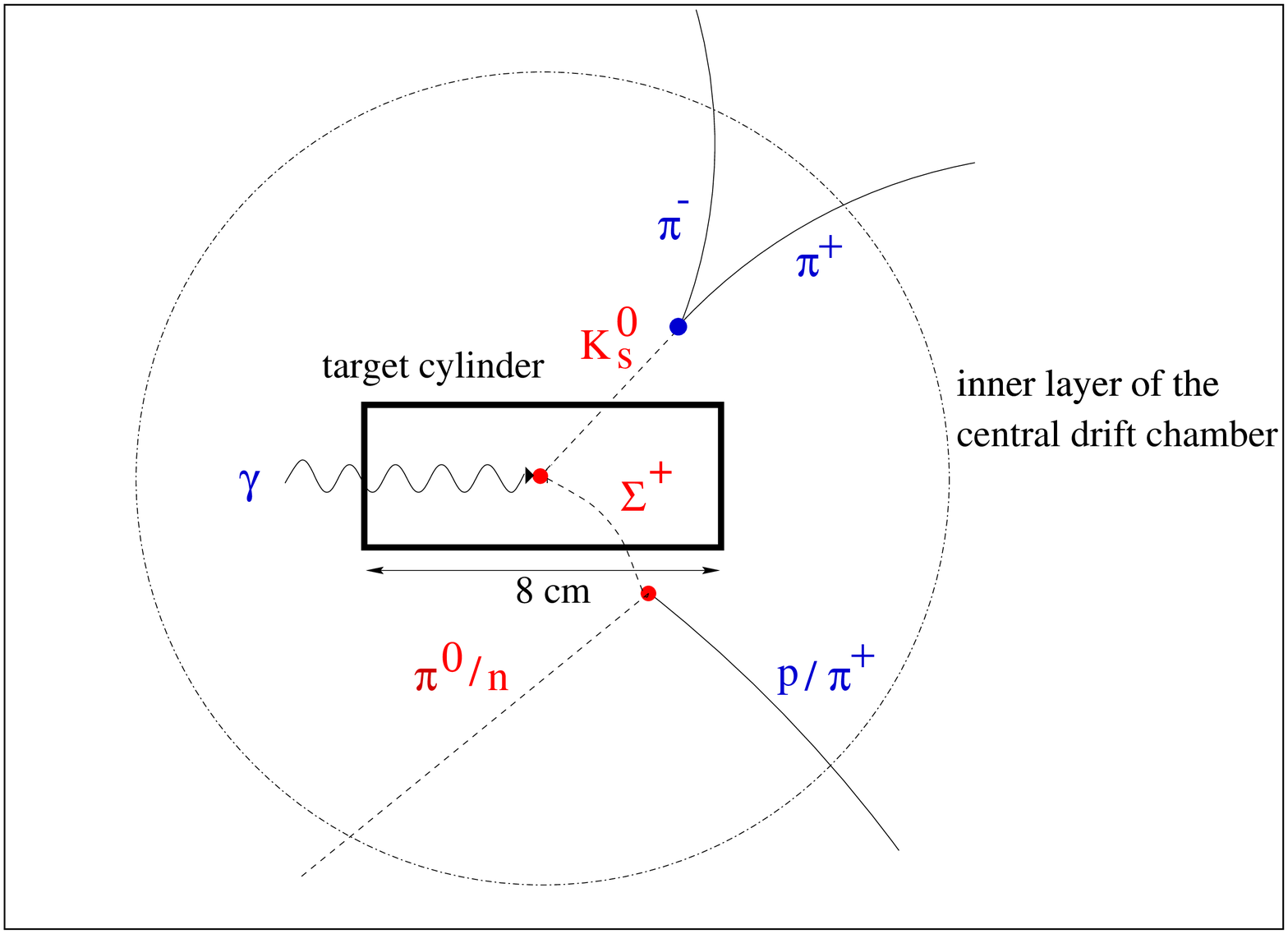}}
\caption{Event topology of $\gamma p \rightarrow K^0_{S} \Sigma^+$ with the decays 
$K^0_{S} \rightarrow \pi^+ \pi^-$ and $\Sigma^+ \rightarrow p \pi^0 / n \pi^+$. 
The full lines indicate the measured tracks of charged particles. The dashed lines 
indicate the unmeasured particle tracks of $K^0_{S}$, $\Sigma^+$ and $n$ or $\pi^0$, 
respectively.}
\label{pic:datevas_info}
\end{figure}

\begin{figure}[h]
\hspace{0.0cm} 
\resizebox{0.5\textwidth}{6.5cm}{\includegraphics{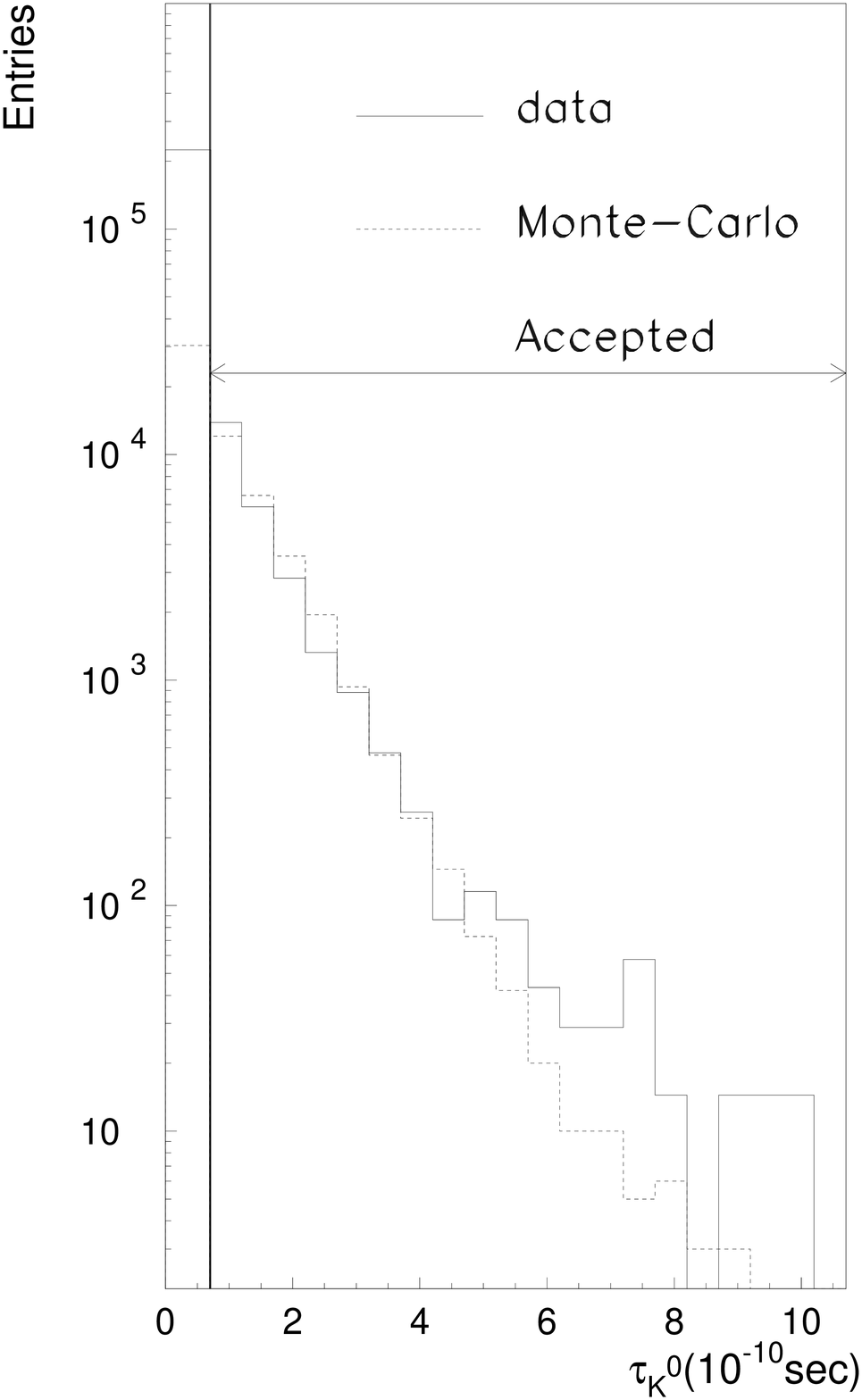}}
    \caption{Decay time distribution of $K^0_{S}$ for data (full line) and for Monte-Carlo simulated 
events (dashed line). The vertical line indicates the cut applied.}
    \label{pic:k0_lifetime}
\hspace{0.0cm}
\resizebox{0.5\textwidth}{6.5cm}{\includegraphics{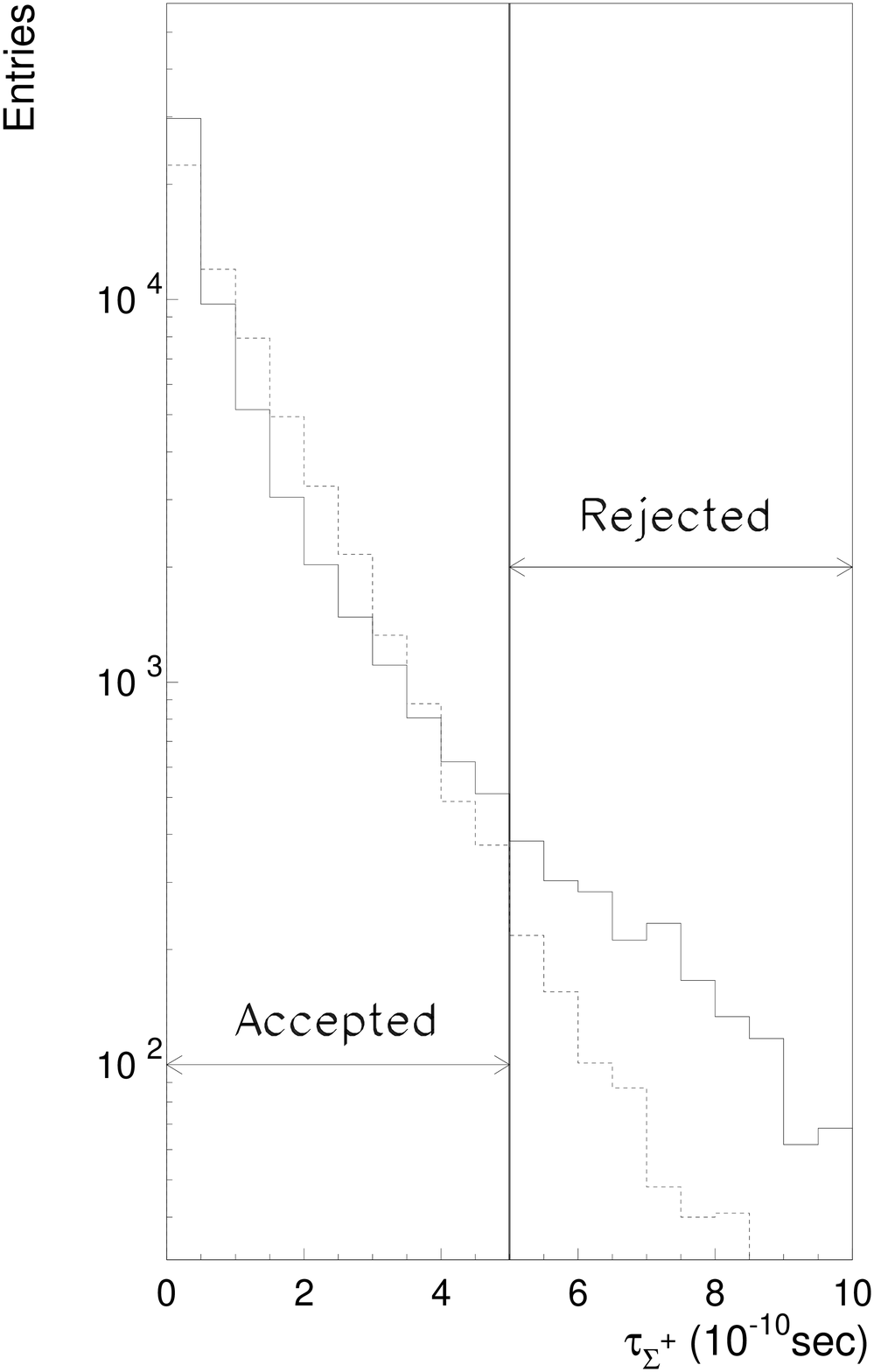}}
    \caption{Decay time distribution of $\Sigma^+$ candidates for data (full line) 
and for Monte-Carlo simulated events (dashed line).  The vertical line indicates the cut applied.}
    \label{pic:sigma_lifetime}
\end{figure}

\begin{figure}[h!]
\hspace{0.0cm}
\resizebox{0.5\textwidth}{6.5cm}{\includegraphics{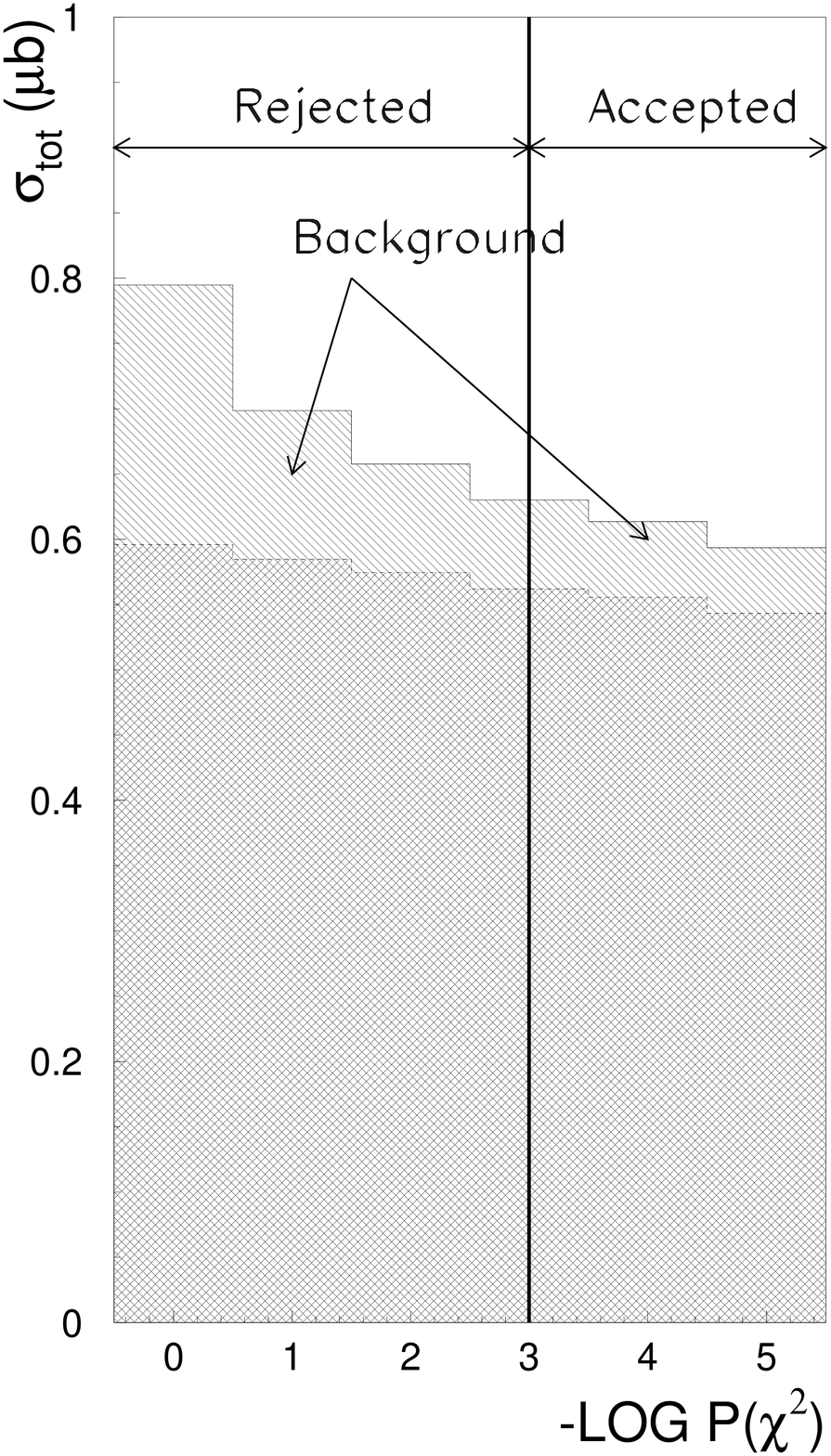}}
\caption{The directly measured reaction cross section and 
the background subtracted cross section (dark grey area) as a function of the cut 
on the primary vertex probability. The light grey area represents the 
background contribution (see text). The vertical line indicates the cut applied.}
\label{pic:lindemannscan}
\end{figure}

The events of the reaction $\gamma p \rightarrow K^0 \Sigma^+$ were reconstructed 
from the measurement of the charged decay products from $K^0_{S} \rightarrow \pi^+ \pi^-$, 
$\Sigma^+ \rightarrow p \pi^0$ and $\Sigma^+ \rightarrow n \pi^+$ respectively (see 
fig. \ref{pic:datevas_info}).
The measured 3-momenta of the positively and negatively charged particles together 
with the incident photon energy (measured in the tagging system) provided all the information 
needed to reconstruct the complete event topology and to determine the 3-momenta of 
the unmeasured particles, i.e. $\Sigma^+$, $K^0_{S}$ and the neutral decay particle 
$\pi^0$ or $n$, respectively, by kinematical fits.\\
First the unmeasured $K^0_{S}$ decay was reconstructed by extrapolating two tracks of 
opposite charge back to the decay vertex. Using the energy 
measurement of the incident 
photon and the reconstructed 3-momentum of the $K^0_{S}$ candidate a kinematical fit 
was carried out assuming the primary reaction $\gamma p \rightarrow K^0_{S}\Sigma^+$ 
which yielded the momentum of the \mbox{missing $\Sigma^+$.}\\
In the next step the primary vertex and the $\Sigma^+$ decay were reconstructed 
simultaneously by an iterative method that was already described in a previous 
publication \cite {Goers99}. The positively charged track 
(not belonging to the $K^0_{S}$ candidate) and the $\Sigma^+$ track at the $\Sigma^+$ decay vertex 
allowed to reconstruct the four-momentum of the neutral decay particle $n$ or $\pi^0$, 
respectively.\\
An event was accepted if it was successfully fitted to the kinematics of the complete 
reaction while the assigned charged particle masses were consistent with the TOF-measurements 
or undetermined (for tracks which were outside the geometrical acceptance of the hodoscopes 
or had no signal due to hodoscope inefficiencies). \\
The reconstructed decay time distributions of $K^0_{S}$ and $\Sigma^+$ are shown in 
figs. \ref{pic:k0_lifetime} and \ref{pic:sigma_lifetime} respectively. 
Monte-Carlo simulated events assuming the proper $K^0_{S}$ lifetime are also drawn.
The comparison indicates that there is a substantial excess of background 
from other reactions mainly at low decay times for $K^0_{S}$ and at a 
smaller scale at large decay times for $\Sigma^+$. In order to reduce this background 
the event sample was restricted to $\tau_{K^0_{S}} > 0.7 \cdot 10^{-10} s$ and 
$\tau_{\Sigma^+} < 5 \cdot 10^{-10} s$.\\
These cuts were not sufficient to remove the falsely included events completely. In 
particular, the main background reactions $\gamma p \rightarrow p\pi^+\pi^-\pi^0$ and 
$n\pi^+\pi^+\pi^-$ were still present in the accepted ranges of the decay times of 
$K^0_{S}$ and $\Sigma^+$ according to the limited spatial resolution of the decay 
vertices. \\
Since such background events must have a common vertex formed by three charged tracks, 
corresponding background reactions were simulated (see sect. \ref{sec:background}) 
and studied with respect to the probability of the vertex fit.
The dependence of the reaction cross section 
from a cut on this probability after all other selection cuts is displayed 
in fig. \ref{pic:lindemannscan} (total distribution). The light grey area 
represents the estimated total background 
from other reactions (see table \ref{tab:bg_beitraege} in sect. \ref{sec:background}). 
Events with a vertex fit probability bigger than $10^{-3}$ were removed both in 
data and simulated background events. 
The cut was chosen so that 
the background contribution to the cross section was at most about $10\%$ on average. 
The final cross section (dark grey area) was obtained by subtraction of 
the accumulated background events (see sect. \ref{results}).

\begin{figure}[ht!]
\resizebox{0.5\textwidth}{5cm}{\includegraphics{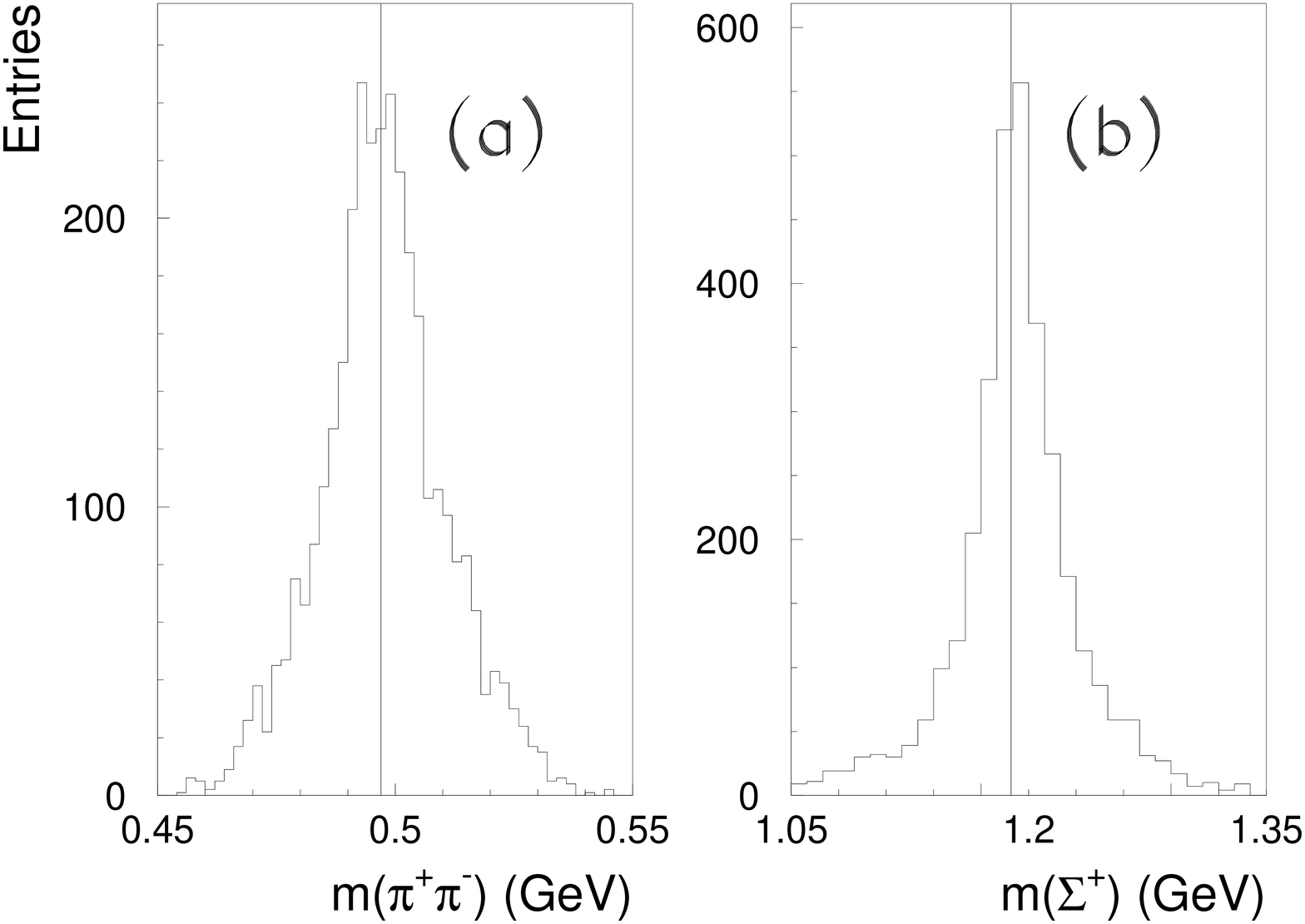}}
    \caption{(a): Distribution of the invariant mass of the $\pi^+ \pi^-$ system;  
(b): Distribution of the missing mass calculated at the primary vertex (see text).}
    \label{pic:minv_k0_miss_si}
\resizebox{0.5\textwidth}{5cm}{ \includegraphics{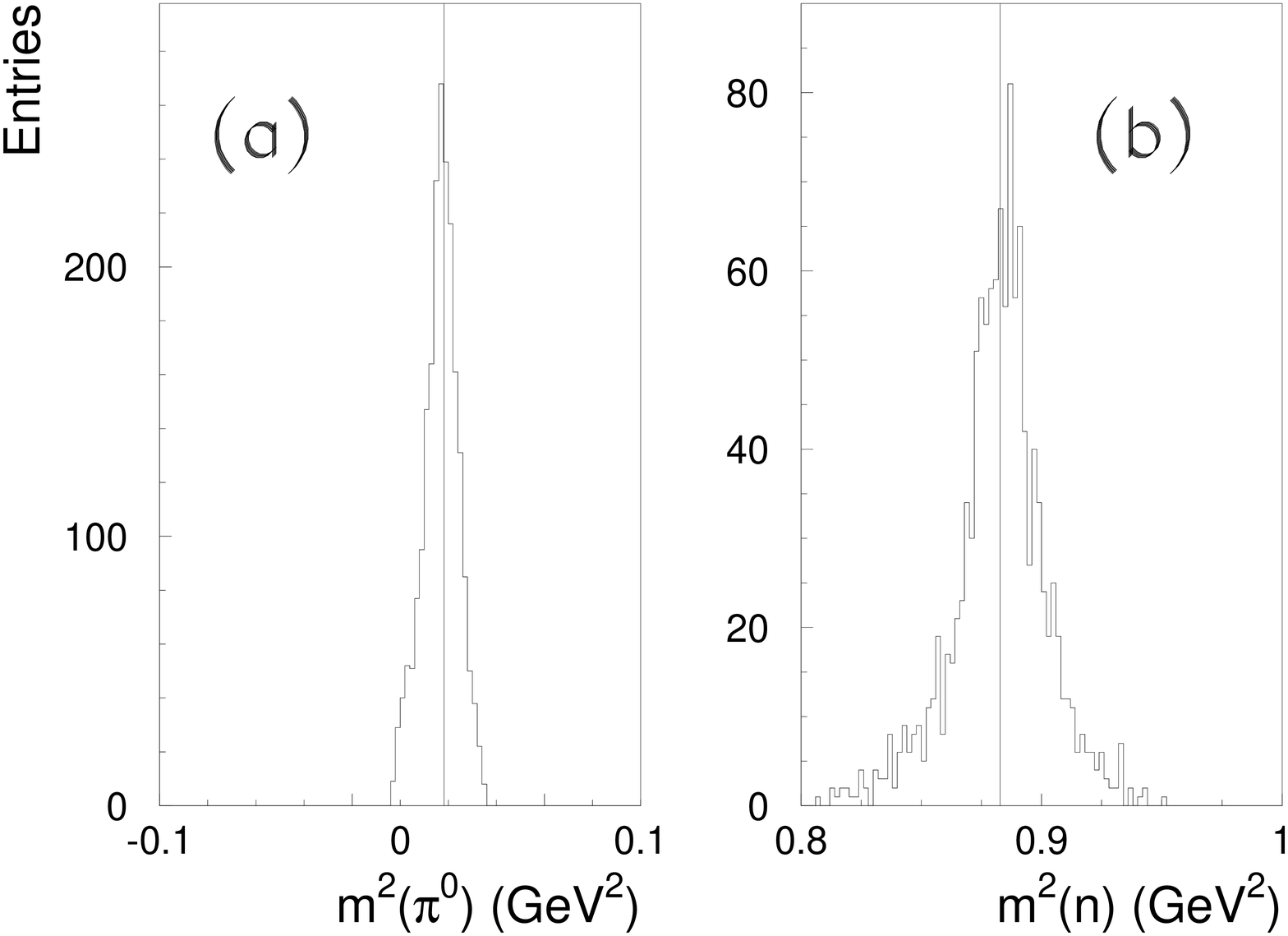}}
    \caption{Distribution of the missing mass squared for \newline (a) $\Sigma^+ \rightarrow p \pi^0$, 
(b) for $\Sigma^+ \rightarrow n \pi^+$ (see text).}
    \label{pic:miss_nfit2}
\end{figure}

The following plots display the final data sample after all cuts: fig. \ref{pic:minv_k0_miss_si} 
depicts the invariant mass distribution of the $\pi^+ \pi^-$ system at the kaon decay vertex 
corresponding to the $K^0_{S}$ mass, and the missing mass distribution calculated at the 
primary vertex, corresponding to the unmeasured $\Sigma^+$ signal. In fig. \ref{pic:miss_nfit2} 
the squared missing mass distributions at the $\Sigma^+$ decay vertex are shown, corresponding 
to $\pi^0$ and $n$, respectively. The missing mass at the primary vertex was calculated 
from the 4-momenta of the incident $\gamma$ and the reconstructed $K^0_{S}$ 
assuming the reaction $\gamma p \rightarrow K^0_{S} \Sigma^+$. The missing mass 
squared at the $\Sigma^+ $ decay vertex was calculated from the 4-momenta 
of $\Sigma^+$ and $p$/$\pi^+$, respectively.

\section{Acceptance of the events}
\label{acceptance}

The acceptance was determined by means of Monte-Carlo simulations. Events 
of the reaction $\gamma p \rightarrow K^0_{S} \Sigma^+$ were generated according 
to phase space with propagation of both particles according to their lifetimes and 
the decays $K^0_{S} \rightarrow \pi^+ \pi^-$ and $\Sigma^+ \rightarrow p \pi^0$/$n \pi^+$. 
Charged particles in the final state were tracked through the drift chamber system of the 
SAPHIR detector taking into account the magnetic field and multiple scattering in all 
materials of the detector setup passed by the particles. Simulated events were treated 
like real events. The determination of the acceptance comprised the trigger efficiency 
of the data taking periods, the event reconstruction efficiency and the data reduction 
according to the event selection cuts. The acceptance was $3.4\%$ on average varying 
between $2$ and $8\%$ throughout the kinematical range spanned by the complete $
K^0_{S}$ production angular range in the CMS and the photon energy range of the data.

\section{Background from other reactions}
\label{sec:background}

\begin{figure}[h!]
\resizebox{0.5\textwidth}{7cm}{\includegraphics{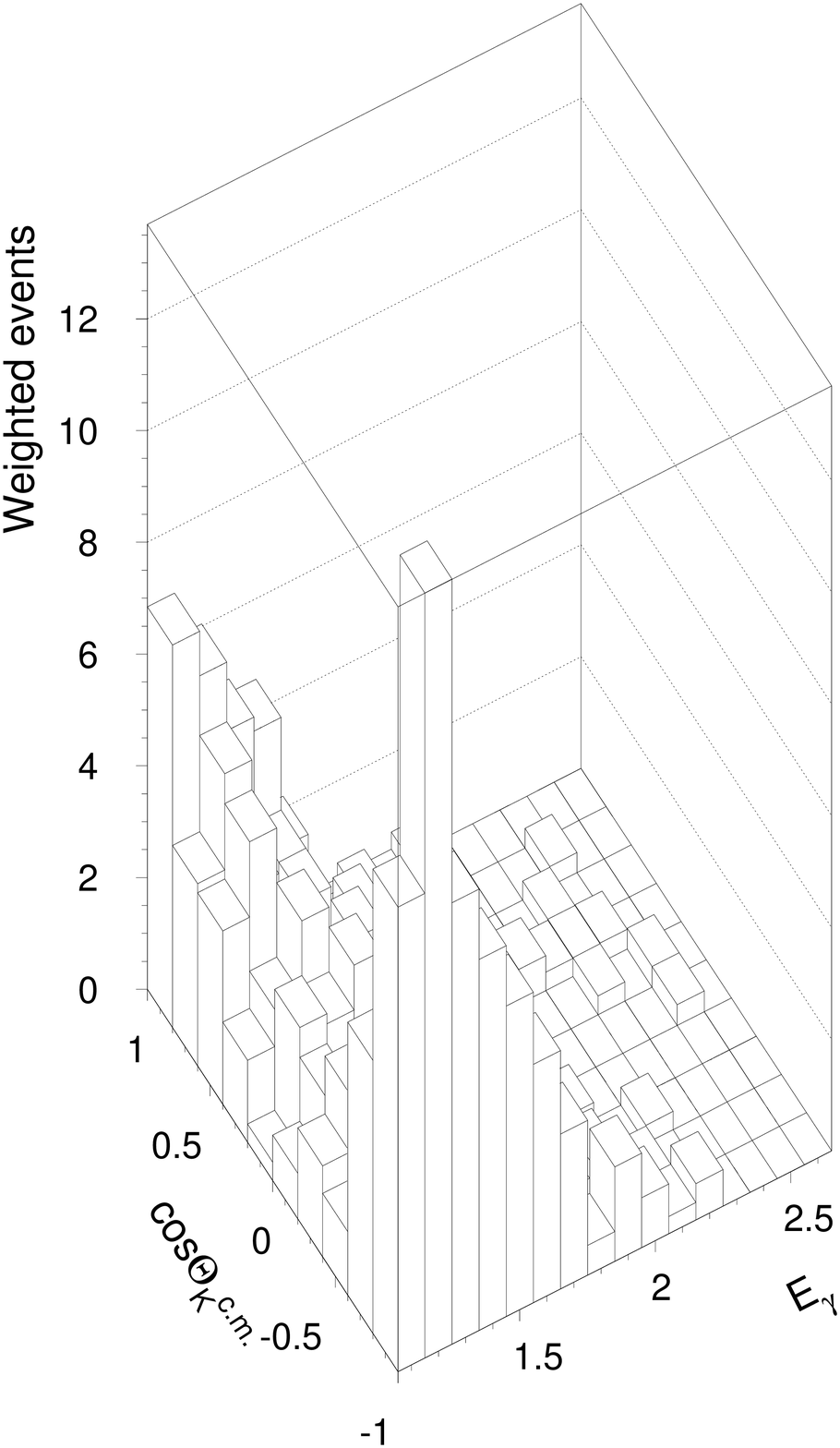}}
    \caption{Distribution of simulated $p\pi^+\pi^-\pi^0$ events as a function 
of the photon energy and the $K^0$ production angle, falsely identified as 
$\gamma p \rightarrow K^0 \Sigma^+$.}
    \label{pic:zaehlrate_bg}
\end{figure}

In order to estimate possible background contributions the reactions listed in 
table \ref{tab:bg_beitraege} were simulated according 
to phase space assuming a $1/E_{\gamma}$ dependence in the photon energy range of the data 
and processed like real data through the reconstruction and selection procedure.\\
The acceptance of background of events falsely selected as $\gamma p \rightarrow K^0_{S}\Sigma^+$ 
and the estimates of their contributions in terms of cross sections obtained on average 
after all cuts are shown in Table \ref{tab:bg_beitraege}. The errors of the cross 
sections (last column) are in the order of $10\%$ or less.\\
The background contributions were also quantified in bins of the photon energy and the $K^0_{S}$ 
production angle. The reactions $\gamma p \rightarrow p \pi^+ \pi^- \pi^0$ and 
$\gamma p \rightarrow n \pi^+ \pi^+ \pi^-$ contribute preferentially at low energy and 
in backward direction in the center of mass system. To give an example, 
fig. \ref{pic:zaehlrate_bg} shows simulated events of $\gamma p \rightarrow p \pi^+ \pi^- \pi^0$ 
which were falsely accepted as events from the reaction $\gamma p \rightarrow K^0 \Sigma^+$ in 
the analysis procedure, distributed in the $E_{\gamma}$ - $\cos\Theta_K^{c.m.}$ plane. The numbers 
of events were weighted by multiplying with ${E_{\gamma}}$ to enhance the 
rare events at high ${E_{\gamma}}$. 
Smaller, but non-negligible contributions stem from the reactions 
$\gamma p \rightarrow K\Lambda\pi$ which contribute preferentially at higher energies and 
towards higher $K^0_{S}$ production angles (not shown).\\
For subtracting the background as a function of the photon energy and the $K^0_{S}$ production 
angle the sum of the contributions of all the reactions was taken into account bin-by-bin 
(see sect. \ref{results}).
\begin{table}[h!]
\caption{Considered reactions, their cross sections throughout the photon energy range, 
calculated acceptances of events falsely selected as reaction $\gamma p \rightarrow K^0 \Sigma^+$ 
($A$), and effective cross sections with which they contribute to $\gamma p \rightarrow K^0 \Sigma^+$ 
(last column). The last row gives the accumulated contributions of all background channels together.}
\label{tab:bg_beitraege} 
\hspace{0.7cm}
\begin{tabular}{lccc}
\hline\noalign{\smallskip}
Reaction & $\sigma$ ($\mu b$) &$A$ ($10^{-2}$) & $\frac{A}{A_{K\Sigma}}$\,$\cdot\,\sigma$ ($ \mu b$) \\
\noalign{\smallskip}\hline\noalign{\smallskip}

$p\pi^+\pi^-$            &  $20$ - $55$ & 0.0002  & 0.0032 \\
$p\pi^+\pi^-\pi^0$       &  $6$ - $28$  & 0.0062  & 0.0503 \\
$n\pi^+\pi^+\pi^-$       &  $3$ - $10$  & 0.0031  & 0.0090 \\
$K^+ \Lambda \pi^0$      &  $ 0.5$    & 0.0100  & 0.0014 \\
$K^0_{S} \Lambda \pi^+$  &  $ 0.5$    & 0.0110  & 0.0016 \\
$K^0_{L} \Lambda \pi^+$  &  $ 0.5$    & 0.0100  & 0.0014 \\
$K^0_{S} \Sigma^+ \pi^0$ &  $ 0.2$    & 0.0060  & 0.0003 \\
$K^+ \Lambda$            &  $ 1.0$    & 0.0030  & 0.0009 \\
$K^+ \Sigma^0$           &  $ 1.0$    & 0.0010  & 0.0003 \\
\noalign{\smallskip}\hline\noalign{\smallskip}
Total                    &            &         & 0.0684 \\
\noalign{\smallskip}\hline
\end{tabular}
\end{table}

\clearpage

\section{Results}
\label{results}

\begin{figure}[b!]
\resizebox{0.55\textwidth}{10cm}{\includegraphics{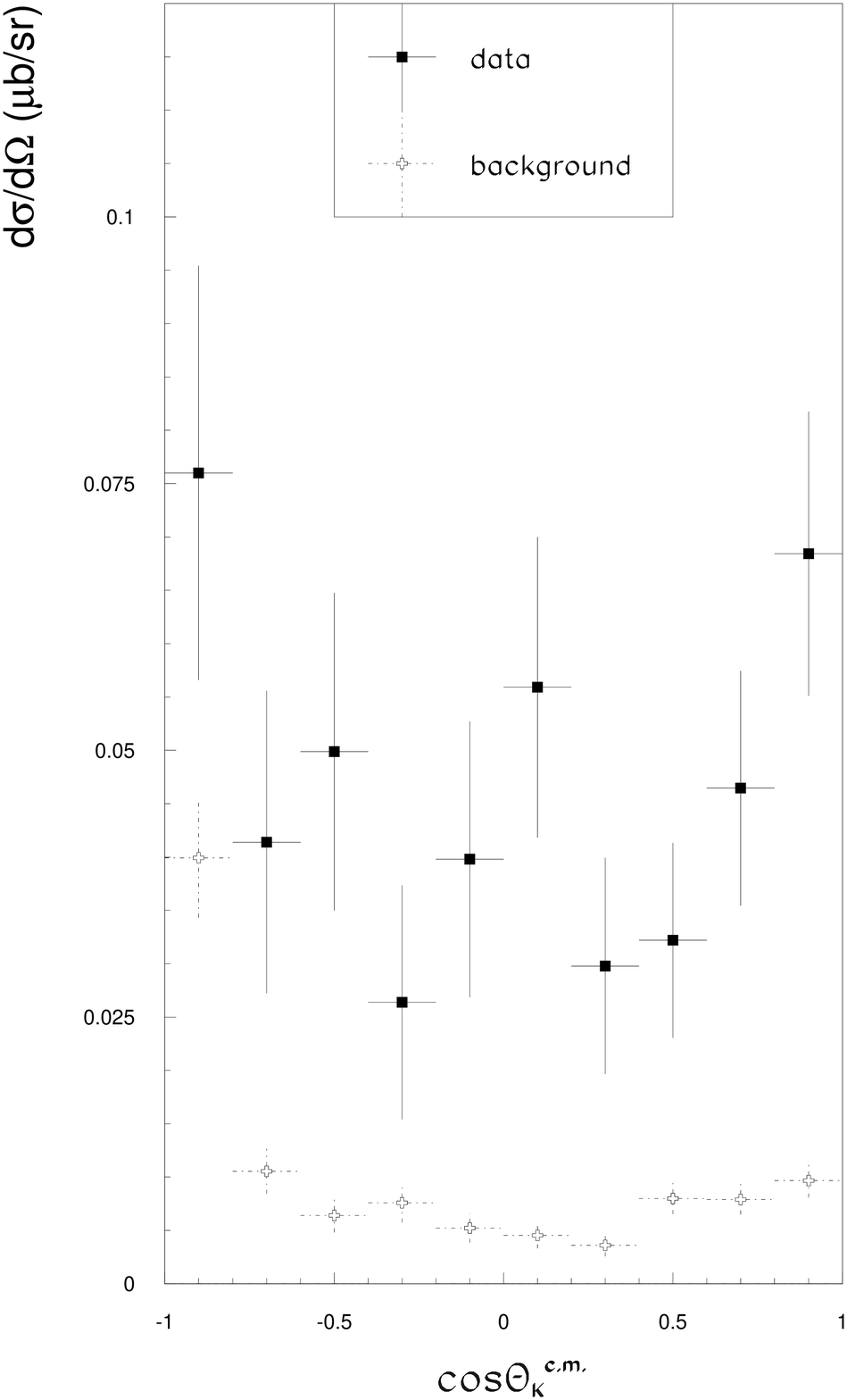}}
    \caption{Example of a measured differential cross section of $\gamma p \rightarrow K^0 \Sigma^+$ 
and the estimation of the total contribution of background from the reactions 
$p\pi^+\pi^-\pi^0$ and $n\pi^+\pi^+\pi^-$. Experimental data are shown in the photon energy range 
between $1.15$ and $1.25$ GeV for one data taking period (february 1998).}
    \label{pic:wq_bg}
\end{figure}

\begin{figure*}[p]
\hspace{1cm}
  \resizebox{0.9\textwidth}{20cm}{\includegraphics{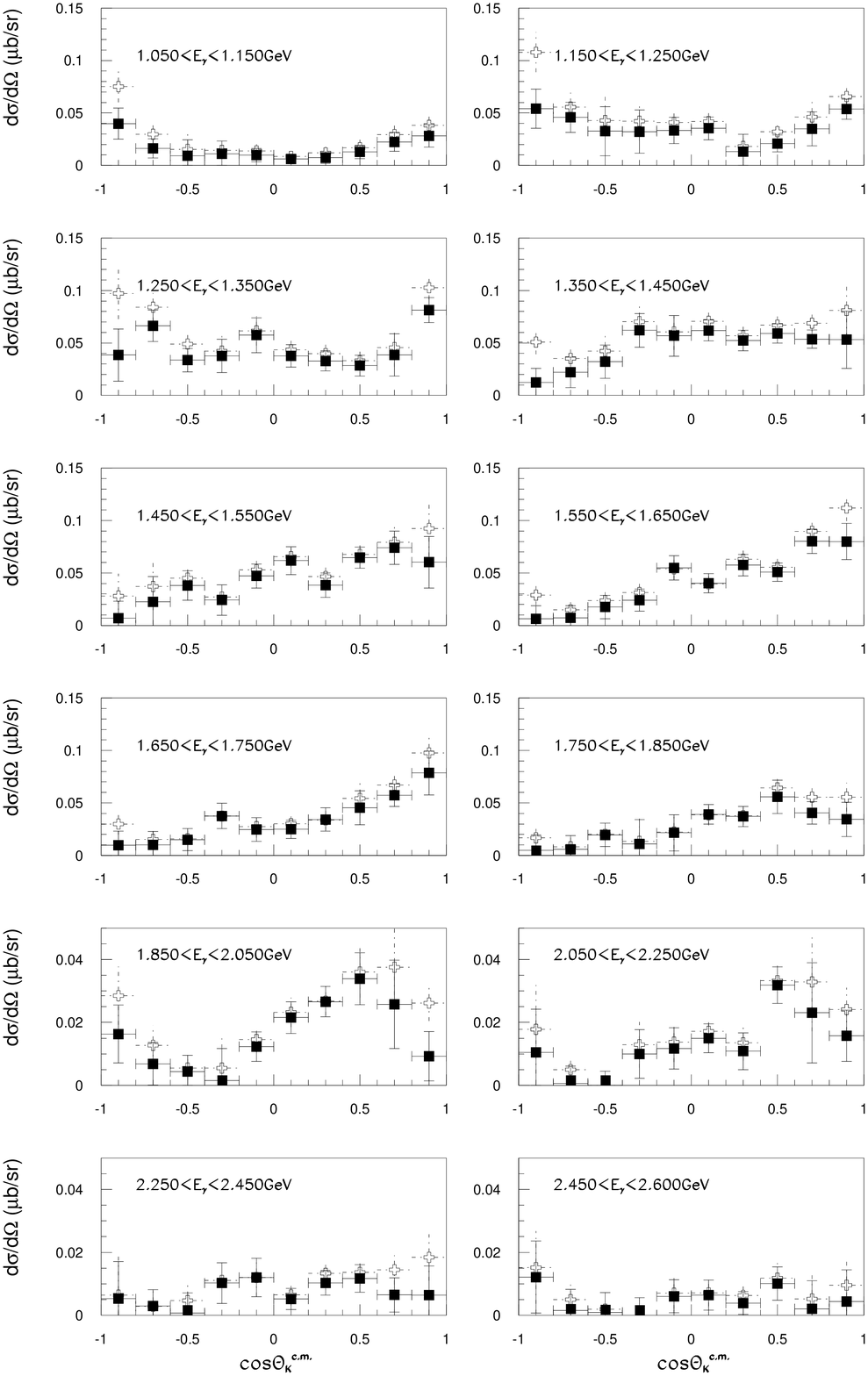}}
    \caption{Differential cross sections $\frac{d\sigma}{d\Omega}$ before (open crosses) and 
after (black squares) subtraction of background.}
    \label{pic:diff_wq}
\end{figure*}

\begin{figure*}[ht!]
\hspace{3cm}
\resizebox{0.6\textwidth}{10cm}{\includegraphics{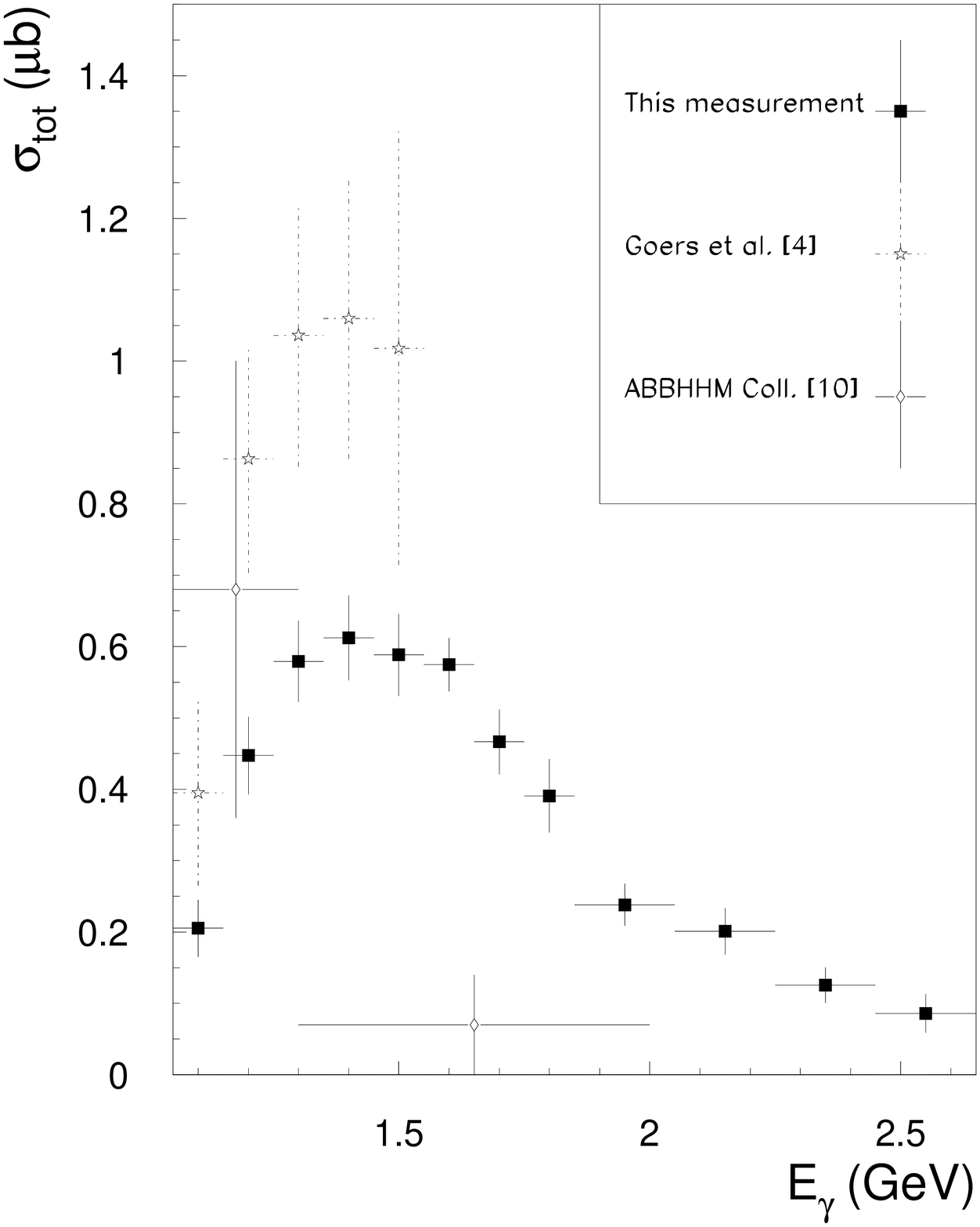}}
    \caption{Total cross section as a function of the photon energy after background 
subtraction in comparison with existing data.}
    \label{pic:tot_wq}
\end{figure*}

\begin{figure*}[hb!]
\hspace{3cm}
\resizebox{0.6\textwidth}{10cm}{\includegraphics{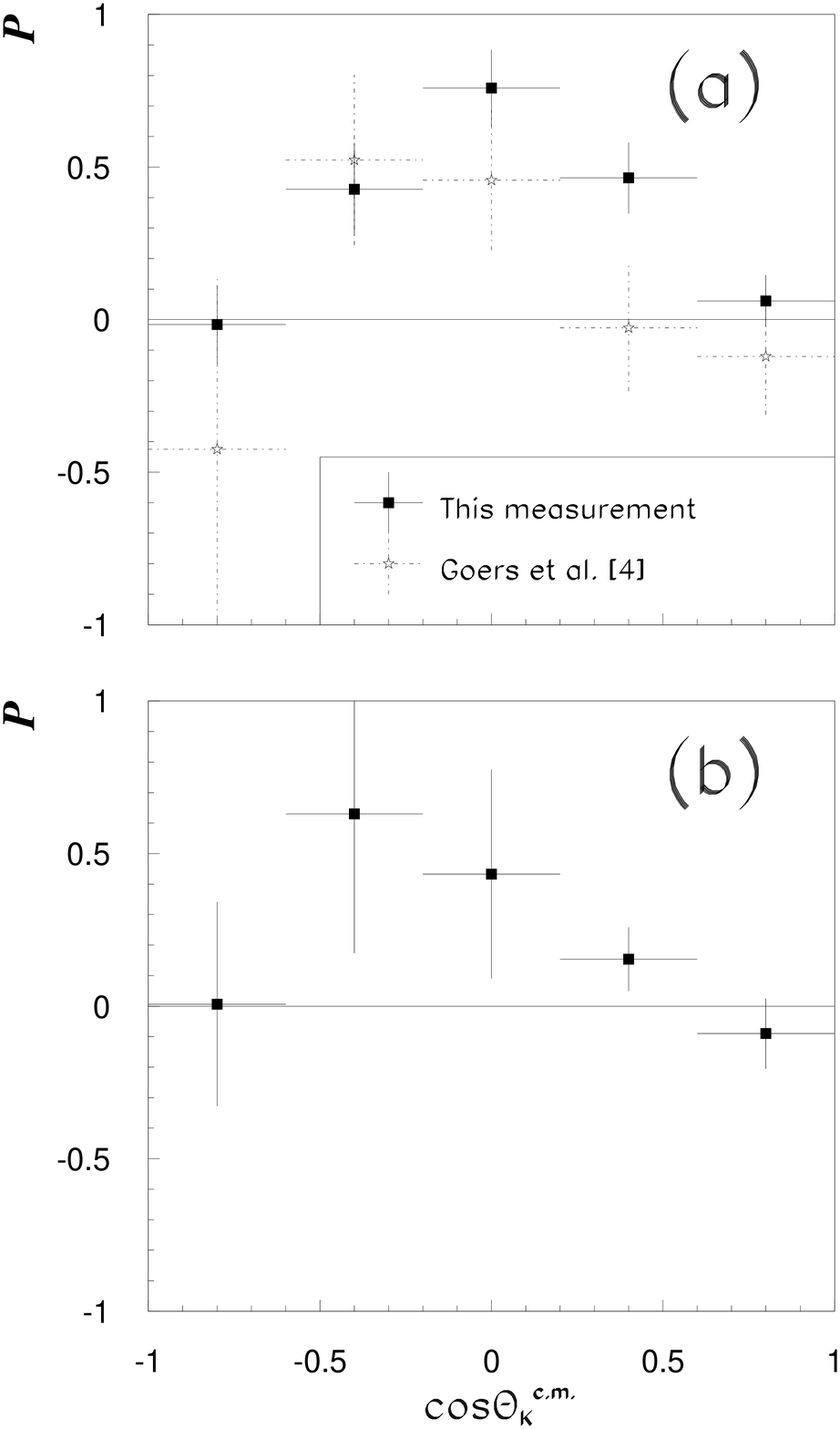}}
    \caption{Polarization parameter of $\Sigma^+$ in five bins of $\cos\Theta_K^{c.m.}$ in the photon energy range 
(a) below $1.55$ GeV and \newline (b) above $1.55$ GeV.}
    \label{pic:pol_low_high}
\end{figure*}

The final event sample of the reaction consisted of 3310 events, 2114 with the 
decay $\Sigma^+ \rightarrow p\pi^0$ and 1196 with the decay $\Sigma^+ \rightarrow n\pi^+$. 
For each data taking period, cross sections were determined in bins of the photon energy, 
$E_{\gamma}$, and of the $K^0$ production angle in the $\gamma p$ center-of-mass system, 
$\cos\Theta_K^{c.m.}$. They were calculated separately for both $\Sigma^+$ decay modes taking 
into account the branching ratios. The cross sections obtained were found to be consistent 
with each other.\\
Background contributions to the cross sections stemming from other reactions 
(see sect. \ref{sec:background}) were also determined separately for each data taking period 
using the same binning. The errors were calculated by quadratic addition of the statistical 
errors and the assumed normalization uncertainty of $10\%$ (see sect. \ref{sec:background}). 
Fig. \ref{pic:wq_bg} shows as an example the distributions of the measured cross section of 
$\gamma p \rightarrow K^0 \Sigma^+$ corrected for the branching ratios of $K^0_{S}$ and 
$\Sigma^+$ decays and the estimated background from 
$\gamma p \rightarrow p\pi^+\pi^-\pi^0$ and $n\pi^+\pi^+\pi^-$ in the photon energy range 
between $1.150$ and $1.250$ GeV.\\
The accumulated background contributions were subtracted bin-by-bin separately for each data 
taking period. The errors of the subtracted cross sections were calculated by quadratic addition 
of the single errors.
Finally, the cross sections were calculated as statistically weighted means of the subtracted 
cross sections of the four data taking periods. Two sorts of errors were calculated: 
at first the statistical error of the weighted mean of the four measurements, 
$\sigma_{w}$, and secondly the standard deviation of the four measurements from the weighted 
mean, $\sigma_{d}$. In case $\sigma_{d} > \sigma_{w}$ it was assumed that the error $\sigma_{d}$ 
accounts for systematic uncertainties in the run conditions which were not properly taken into 
account in the simulation. The larger of the two errors was accepted as total error. A more 
detailed discussion of the error calculation can be found in \cite{Glan03a}, \cite{Lawall}.\\
Differential cross sections after subtraction are shown in fig.\ref{pic:diff_wq}. The 
corresponding values are given in table \ref{tab:diff_wq}. \\
The reaction cross sections as a function of the photon energy were obtained by 
summing up the differential cross sections over the angular range. Errors were determined 
by quadratic addition of the single errors in the angular bins. The results are shown in 
fig. \ref{pic:tot_wq} and table \ref{tab:tot_wq}. Previous measurements \cite{Goers99}, \cite{ABBHHM69} 
are also shown.\\
The cross sections from the previous data set taken at SAPHIR \cite{Goers99} are 
$30$ to $40\%$ higher than the current results. Careful investigations carried 
out in the course of the new analysis have shown that background, in particular from 
$\gamma p \rightarrow p\pi^+\pi^-\pi^0$, had not been removed sufficiently in the analysis 
of the first data set and moreover the error calculations did not account sufficiently 
for this uncertainty.\\
The polarization of the $\Sigma^+$-hyperon has been measured through its parity-violating 
weak decay. The angular distribution of the decay nucleon is given by \cite{LeeGatto}: 
\begin{displaymath}
W(\theta_{N}) = \frac{1}{2} \cdot (1\,+\,\alpha\,\cdot\,P\,\cdot\,\cos(\theta_{N}))
\end{displaymath} 
where $\alpha$ is the asymmetry parameter which is equal to $(-0.98^{+0.017}_{-0.015})$ for 
$\Sigma^+ \rightarrow p\pi^0$ and $(0.068^{+0.013}_{-0.013})$ for \mbox{$\Sigma^+ \rightarrow n\pi^+$} 
\cite{PDG}. The parameter P measures the hyperon polarization and $\theta_{N}$ is the decay 
angle of the nucleon measured with respect to the normal to the production plane of 
$K^0_{S}$ and $\Sigma^+$ in the $\Sigma^+$ rest system.\\
P has been determined by fits to the polarization angular distribution for the decay mode 
$\Sigma^+ \rightarrow p\pi^0$ as a function of $\cos\Theta_K^{c.m.}$ for photon energies 
below and above $1.55$ GeV. This decay mode is more sensitive  than $\Sigma^+ \rightarrow n\pi^+$ 
due to the larger asymmetry parameter. The values of $P$ are shown in fig. \ref{pic:pol_low_high} 
and table \ref{tab:pol} together with the previous measurement at SAPHIR \cite{Goers99}. 
The error bars refer to $\sigma_{w}$ and $\sigma_{d}$ respectively as described above.

\clearpage
\newpage

\begin{table*}
\caption{Differential cross sections $d\sigma/d\Omega$ in $\mu b$ in 12 bins of $E_{\gamma}$ and 
10 bins of $\cos\Theta_K^{c.m.}$, obtained after background subtraction.\vspace{1.0cm}}
\label{tab:diff_wq}       % Give a unique label
\hspace{3.0cm}
\vspace{1.0cm}
\begin{tabular}{lllll}
\hline\noalign{\smallskip}
              $\cos\Theta_K^{c.m.}$  & \multicolumn{4}{c}{$E_{\gamma}$ (GeV)} \\ \hline
                    & 1.050 - 1.150 & 1.150 - 1.250 & 1.250 - 1.350 & 1.350 - 1.450 \\ \hline

\noalign{\smallskip}\hline\noalign{\smallskip}

-1.0 $-$  -0.8  &  0.040 $\pm$ 0.015  &  0.054 $\pm$ 0.019  &  0.038 $\pm$ 0.025  &  0.012 $\pm$ 0.013  \\ \hline 
-0.8 $-$  -0.6  &  0.016 $\pm$ 0.009  &  0.046 $\pm$ 0.014  &  0.066 $\pm$ 0.015  &  0.022 $\pm$ 0.014  \\ \hline 
-0.6 $-$  -0.4  &  0.009 $\pm$ 0.015  &  0.033 $\pm$ 0.023  &  0.034 $\pm$ 0.011  &  0.032 $\pm$ 0.016  \\ \hline 
-0.4 $-$  -0.2  &  0.011 $\pm$ 0.012  &  0.032 $\pm$ 0.020  &  0.038 $\pm$ 0.016  &  0.062 $\pm$ 0.016  \\ \hline 
-0.2 $-$ \,\,0.0 &  0.010 $\pm$ 0.006  &  0.033 $\pm$ 0.013  &  0.057 $\pm$ 0.017  &  0.057 $\pm$ 0.020  \\ \hline 
\,\,0.0 $-$ \,\,0.2 &  0.006 $\pm$ 0.005  &  0.035 $\pm$ 0.011  &  0.038 $\pm$ 0.011  &  0.062 $\pm$ 0.010  \\ \hline 
\,\,0.2 $-$ \,\,0.4 &  0.007 $\pm$ 0.006  &  0.013 $\pm$ 0.016  &  0.033 $\pm$ 0.010  &  0.052 $\pm$ 0.010  \\ \hline 
\,\,0.4 $-$ \,\,0.6 &  0.013 $\pm$ 0.006  &  0.021 $\pm$ 0.008  &  0.028 $\pm$ 0.010  &  0.059 $\pm$ 0.009  \\ \hline 
\,\,0.6 $-$ \,\,0.8 &  0.022 $\pm$ 0.009  &  0.035 $\pm$ 0.016  &  0.039 $\pm$ 0.020  &  0.054 $\pm$ 0.009  \\ \hline 
\,\,0.8 $-$ \,\,1.0 &  0.028 $\pm$ 0.011  &  0.054 $\pm$ 0.010  &  0.081 $\pm$ 0.012  &  0.053 $\pm$ 0.027  \\ \hline 

\noalign{\smallskip}\hline
\end{tabular}
\end{table*}

\begin{table*}
\label{tab:2}       % Give a unique label
\hspace{3.0cm}
\vspace{1cm}
\begin{tabular}{lllll}
\hline\noalign{\smallskip}

              $\cos\Theta_K^{c.m.}$  & \multicolumn{4}{c}{$E_{\gamma}$ (GeV)} \\ \hline
                    & 1.450 - 1.550 & 1.550 - 1.650 & 1.650 - 1.750 & 1.750 - 1.850 \\ \hline

-1.0 $-$ -0.8 &  0.007 $\pm$ 0.016  &  0.006 $\pm$ 0.013  &  0.010 $\pm$ 0.013  &  0.005 $\pm$ 0.015  \\ \hline 
-0.8 $-$ -0.6 &  0.022 $\pm$ 0.024  &  0.007 $\pm$ 0.012  &  0.010 $\pm$ 0.013  &  0.006 $\pm$ 0.013  \\ \hline 
-0.6 $-$ -0.4 &  0.038 $\pm$ 0.014  &  0.017 $\pm$ 0.011  &  0.015 $\pm$ 0.011  &  0.020 $\pm$ 0.011  \\ \hline 
-0.4 $-$ -0.2 &  0.024 $\pm$ 0.015  &  0.024 $\pm$ 0.011  &  0.038 $\pm$ 0.012  &  0.011 $\pm$ 0.023  \\ \hline 
-0.2 $-$ \,\,0.0 &  0.047 $\pm$ 0.011  &  0.055 $\pm$ 0.012  &  0.025 $\pm$ 0.011  &  0.021 $\pm$ 0.017  \\ \hline 
 \,\,0.0 $-$ \,\,0.2 &  0.062 $\pm$ 0.013  &  0.040 $\pm$ 0.009  &  0.025 $\pm$ 0.009  &  0.039 $\pm$ 0.009  \\ \hline 
\,\,0.2 $-$ \,\,0.4 &  0.038 $\pm$ 0.012  &  0.057 $\pm$ 0.010  &  0.034 $\pm$ 0.011  &  0.037 $\pm$ 0.010  \\ \hline 
\,\,0.4 $-$ \,\,0.6 &  0.065 $\pm$ 0.010  &  0.051 $\pm$ 0.009  &  0.045 $\pm$ 0.016  &  0.056 $\pm$ 0.016  \\ \hline 
\,\,0.6 $-$ \,\,0.8 &  0.074 $\pm$ 0.016  &  0.080 $\pm$ 0.012  &  0.057 $\pm$ 0.011  &  0.040 $\pm$ 0.011  \\ \hline 
\,\,0.8 $-$ \,\,1.0 &  0.060 $\pm$ 0.024  &  0.080 $\pm$ 0.017  &  0.079 $\pm$ 0.021  &  0.034 $\pm$ 0.016  \\ \hline 

\noalign{\smallskip}\hline
\end{tabular}
\end{table*}
\begin{table*}
\label{tab:3}       % Give a unique label
\hspace{3.0cm}
\vspace{1cm}
\begin{tabular}{lllll}
\hline\noalign{\smallskip}

              $\cos\Theta_K^{c.m.}$  & \multicolumn{4}{c}{$E_{\gamma}$ (GeV)} \\ \hline
                    & 1.850 - 2.050 & 2.050 - 2.250 & 2.250 - 2.450 & 2.450 - 2.600 \\ \hline

-1.0 $-$ -0.8 &  0.016 $\pm$ 0.009  &  0.010 $\pm$ 0.014  &  0.005 $\pm$ 0.012  &  0.012 $\pm$ 0.011  \\ \hline 
-0.8 $-$ -0.6 &  0.007 $\pm$ 0.007  &  0.001 $\pm$ 0.005  &  0.003 $\pm$ 0.005  &  0.002 $\pm$ 0.007  \\ \hline 
-0.6 $-$ -0.4 &  0.004 $\pm$ 0.005  &  0.000 $\pm$ 0.005  &  0.001 $\pm$ 0.006  &  0.001 $\pm$ 0.006  \\ \hline 
-0.4 $-$ -0.2 &  0.002 $\pm$ 0.010  &  0.010 $\pm$ 0.008  &  0.010 $\pm$ 0.007  &  0.000 $\pm$ 0.006  \\ \hline 
-0.2 $-$ \,\,0.0 &  0.012 $\pm$ 0.005  &  0.012 $\pm$ 0.007  &  0.012 $\pm$ 0.006  &  0.006 $\pm$ 0.005  \\ \hline 
\,\,0.0 $-$ \,\,0.2 &  0.022 $\pm$ 0.005  &  0.015 $\pm$ 0.005  &  0.005 $\pm$ 0.003  &  0.006 $\pm$ 0.005  \\ \hline 
\,\,0.2 $-$ \,\,0.4 &  0.027 $\pm$ 0.005  &  0.011 $\pm$ 0.006  &  0.010 $\pm$ 0.004  &  0.004 $\pm$ 0.004  \\ \hline 
\,\,0.4 $-$ \,\,0.6 &  0.034 $\pm$ 0.008  &  0.032 $\pm$ 0.006  &  0.012 $\pm$ 0.004  &  0.010 $\pm$ 0.005  \\ \hline 
\,\,0.6 $-$ \,\,0.8 &  0.026 $\pm$ 0.014  &  0.023 $\pm$ 0.016  &  0.007 $\pm$ 0.005  &  0.002 $\pm$ 0.009  \\ \hline 
\,\,0.8 $-$ \,\,1.0 &  0.009 $\pm$ 0.008  &  0.016 $\pm$ 0.008  &  0.006 $\pm$ 0.009  &  0.004 $\pm$ 0.010  \\ \hline 

\noalign{\smallskip}\hline
\end{tabular}
\end{table*}

\clearpage

\begin{table*}
\caption{Total reaction cross sections in 12 bins of $E_{\gamma}$, obtained after background subtraction.\vspace{1.0cm}}
\hspace{1cm}
\label{tab:tot_wq}       % Give a unique label
\hspace{5.0cm}
\vspace{1cm}
\begin{tabular}{ll}
\hline\noalign{\smallskip}
$E_\gamma$ [GeV] & $\sigma_{tot}$ ($\mu b$) \\  \hline
$1.050\,-\,1.150$  & 0.205 $\pm$ 0.040   \\ \hline
$1.150\,-\,1.250$  & 0.447 $\pm$ 0.054   \\ \hline
$1.250\,-\,1.350$  & 0.579 $\pm$ 0.057   \\ \hline
$1.350\,-\,1.450$  & 0.612 $\pm$ 0.059   \\ \hline
$1.450\,-\,1.550$  & 0.588 $\pm$ 0.058   \\ \hline
$1.550\,-\,1.650$  & 0.575 $\pm$ 0.037   \\ \hline
$1.650\,-\,1.750$  & 0.466 $\pm$ 0.045   \\ \hline
$1.750\,-\,1.850$  & 0.391 $\pm$ 0.051   \\ \hline
$1.850\,-\,2.050$  & 0.238 $\pm$ 0.029   \\ \hline
$2.050\,-\,2.250$  & 0.201 $\pm$ 0.033   \\ \hline
$2.250\,-\,2.450$  & 0.125 $\pm$ 0.025   \\ \hline
$2.450\,-\,2.600$  & 0.086 $\pm$ 0.027  \\ \hline
\noalign{\smallskip}\hline
\end{tabular}
\end{table*}
\begin{table*}
\caption{Polarization parameter as a function of $\cos\Theta_K^{c.m.}$ in the photon 
energy range (a) up to $1.55$ GeV and (b) above $1.55$ GeV.\vspace{1.0cm}}
\label{tab:pol}       % Give a unique label
\hspace{5.0cm}
\vspace{1cm}
\begin{tabular}{lll}
\hline\noalign{\smallskip}
 $\cos\Theta_K^{c.m.}$ & $E_{\gamma}< 1.55$ GeV & $E_{\gamma}> 1.55$ GeV \\  \hline
 -1.0 \,$-$\,   -0.6 &     -0.016 $\pm$ 0.128 &  \,\,0.007 $\pm$ 0.334 \\  \hline
 -0.6 \,$-$\,   -0.2 &  \,\,0.428 $\pm$ 0.152 &  \,\,0.630 $\pm$ 0.457 \\  \hline
 -0.2 \,$-$\,\,\,    0.2  &  \,\,0.759 $\pm$ 0.126 &  \,\,0.433 $\pm$ 0.343 \\  \hline
 \,\,0.2 \,$-$\,\,\, 0.6  &  \,\,0.464 $\pm$ 0.116 &  \,\,0.154 $\pm$ 0.105 \\  \hline
 \,\,0.6 \,$-$\,\,\, 1.0  &  \,\,0.061 $\pm$ 0.085 & -0.090 $\pm$ 0.116 \\  \hline

\noalign{\smallskip}\hline
\end{tabular}
\end{table*}

\clearpage

\section{Comparison with $\gamma p \rightarrow K^+\Sigma^0$}
\label{comparison_kplsi0}

\begin{figure*}[h!]
\hspace{3cm}
\resizebox{0.6\textwidth}{10cm}{\includegraphics{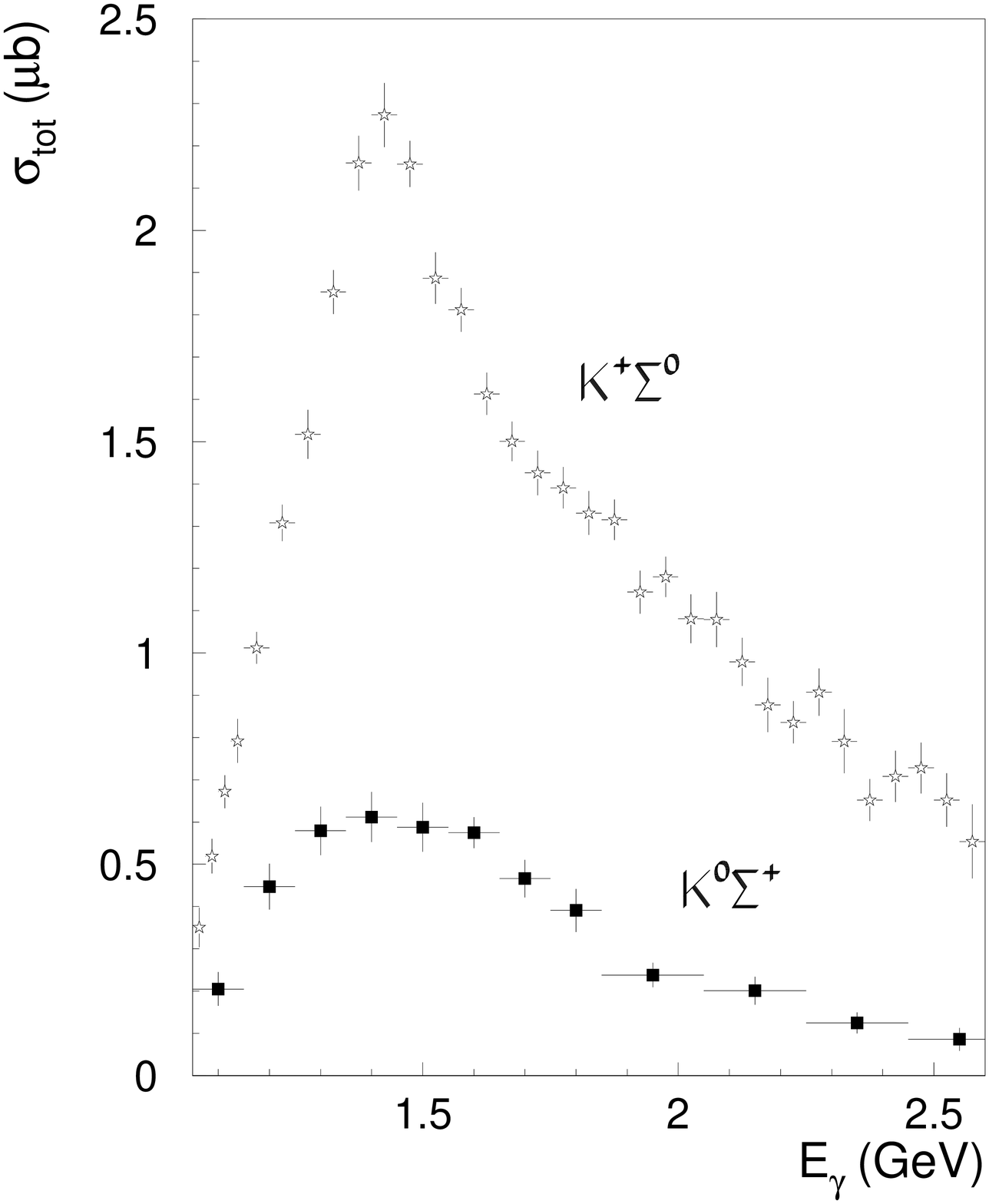}}
    \caption{Comparison of the total cross sections for $\gamma p \rightarrow K^0\Sigma^+$ 
(black dots) and $\gamma p \rightarrow K^+\Sigma^0$ (open stars).}
    \label{pic:totwq_vergleich}
\end{figure*}

\begin{figure*}[h!]
\hspace{3cm}
\resizebox{0.6\textwidth}{10cm}{\includegraphics{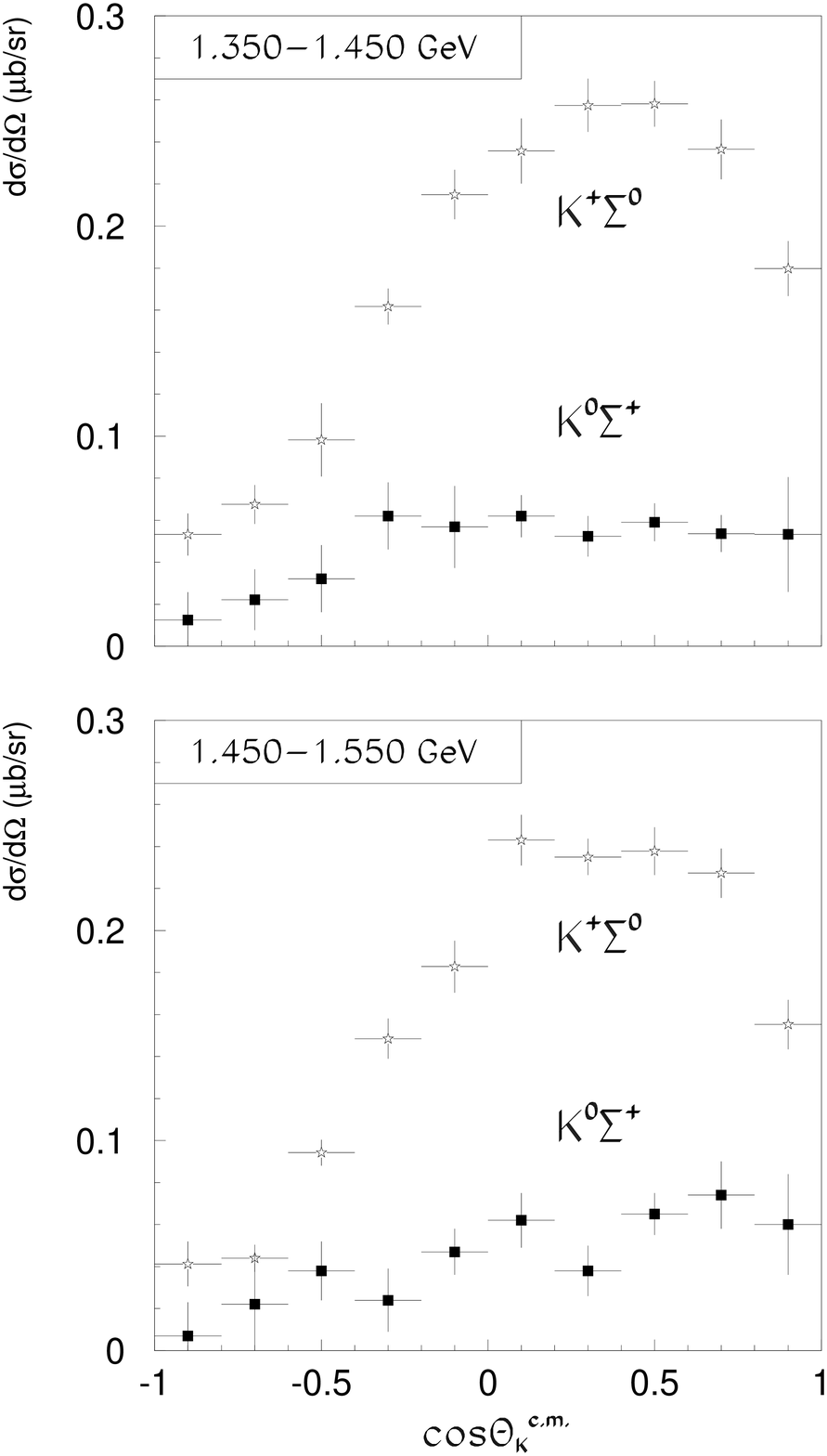}}
    \caption{Comparison of the differential cross sections for $\gamma p \rightarrow K^0\Sigma^+$ 
(black squares) and $\gamma p \rightarrow K^+\Sigma^0$ (open stars).}
    \label{pic:diffwq_vergleich}
\end{figure*}

The reaction $\gamma p \rightarrow K^+\Sigma^0$ was measured in the same experiment 
\cite{Glan03b}. In the following the results are compared to 
$\gamma p \rightarrow K^0\Sigma^+$.\\
Both reaction cross sections culminate at a photon energy around $1.45$ GeV (see 
fig. \ref{pic:totwq_vergleich}). However, while $K^+\Sigma^0$ shows a pronounced peak 
at this energy, $K^0\Sigma^+$ varies more slowly and the cross section stays on a 
much lower level. If the cross sections in the peak region are dominated by the 
production of a $\Delta$ resonance (e.g. $S_{31}(1900)$ or $P_{31}(1910)$) Clebsch-Gordan 
coefficients predict that the ratio of cross sections is four in favor of $K^+\Sigma^0$. 
The difference between both reactions is also evident in the angular dependence. As an 
example, the comparison of the $\frac{d\sigma}{d\Omega}$ distributions in the peak region 
is shown in fig. \ref{pic:diffwq_vergleich}.\\
\begin{figure*}[h!]
\hspace{3cm}
\resizebox{0.6\textwidth}{10cm}{\includegraphics{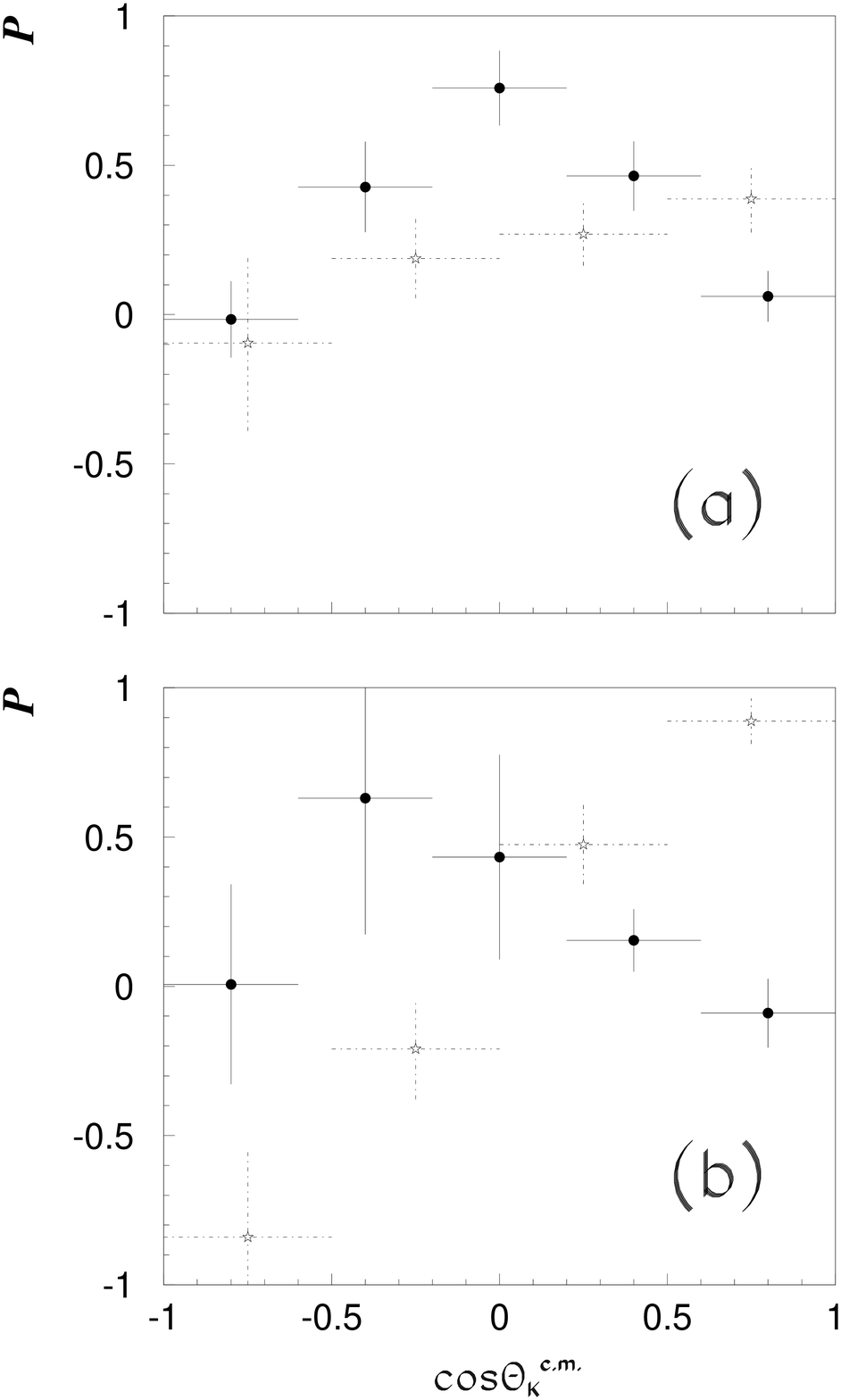}}
    \caption{Comparison of the polarization for $\gamma p \rightarrow K^0\Sigma^+$ 
(black squares) and $\gamma p \rightarrow K^+\Sigma^0$ (open stars) in the photon 
energy range (a) below $1.55$ GeV and (b) above $1.55$ GeV.}
    \label{pic:pol_vergleich}
\end{figure*}
The polarizations of $\Sigma^+$ and $\Sigma^0$ are compared to each other in 
fig. \ref{pic:pol_vergleich}. The $\Sigma^0$ polarization measured in bins 
of $cos \theta_K^{c.m.}$ is found to rise as 
a function of the kaon production angle \cite{Glan03b} while the $\Sigma^+$ 
polarization tends to go through a maximum value around $\cos\theta_K^{c.m.} \approx 0$. 
The $\Sigma^0$ polarization grows in magnitude when going from the lower to the higher energies. 
The $\Sigma^+$ polarization is consistent with little energy dependence.\\
An analysis of the data with respect to the underlying production mechanisms  
and resonance production can be made by the comparison to model calculations \mbox{(see section 
\ref{comparison_theory})}. 
\clearpage

\section{Comparison with theory}
\label{comparison_theory}

\begin{figure}[t]
  \resizebox{0.5\textwidth}{!}{\includegraphics{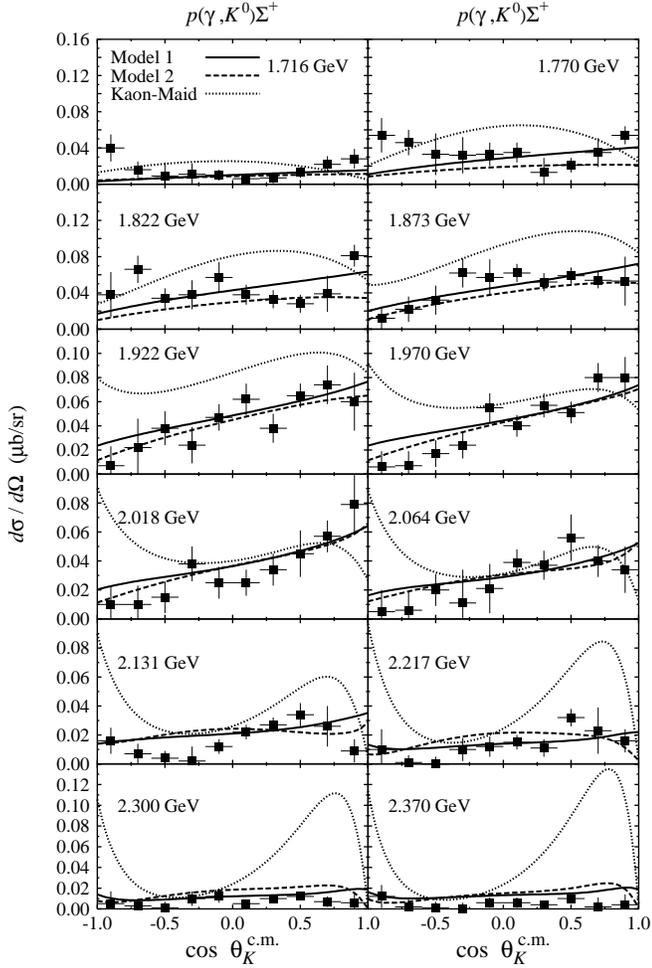}}
    \caption{Differential cross sections for $K^0\Sigma^+$ photoproduction.
      Dotted lines are prediction from the KAON-MAID model \protect\cite{maid},
      dashed lines are obtained from Model 2, whereas solid lines exhibit the  
      prediction of Model 1. The c.m. energy $W$ is shown in each panel.}
    \label{pic:modelsdiffwq}
\end{figure}

\begin{figure}[h]
  \resizebox{0.5\textwidth}{!}{\includegraphics{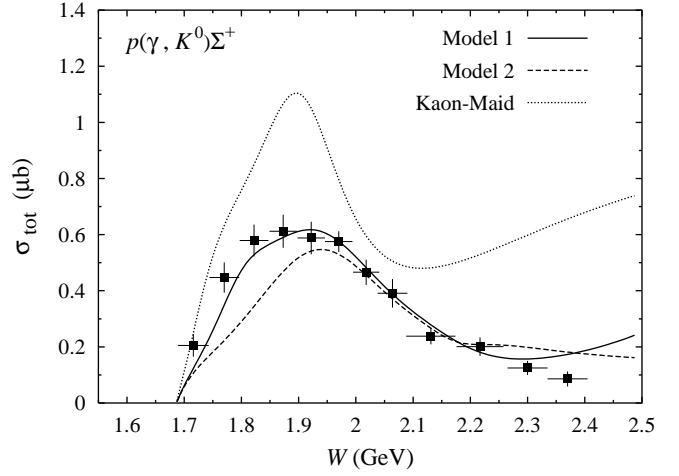}}
    \caption{As in Fig.\ \ref{pic:modelsdiffwq}, but for the total cross section.}
    \label{pic:modelswqtot}
\end{figure}

Here we discuss the comparison with model calculations of Bennhold and Mart \cite{bennhold} 
as well as with previous results from KAON-MAID \cite{maid}. 
Since the phenomenological KAON-MAID 
model\footnote{The model consists of a background part which is constructed from
the standard Born terms, $K^*$ and $K_1$ $t$-channel exchanges, and the resonance
part which includes the isospin 1/2 nucleon resonances $S_{11}(1650)$, $P_{11}(1710)$,
and $P_{13}(1720)$, as well as the isospin 3/2 deltas $S_{31}(1900)$ and
$P_{31}(1910)$.} \cite{bennhold,maid} was fitted to previous SAPHIR data
\cite{Goers99,saphir98}, it obviously overpredicts the cross sections of this new
measurement. This is clearly elucidated in Figs. \ref{pic:modelsdiffwq} 
and \ref{pic:modelswqtot}. In order to see the effect of new data on
the model, we refitted the corresponding coupling constants to the latest
$K^+\Sigma^0$ data \cite{Glan03b} together with the $K^0\Sigma^+$ data presented here using fixed values 
of the resonance parameters \cite{PDG}.  
The result is shown in columns (b) of Table \ref{tab:coupling_constant}. For comparison, 
column (a) contains the result of the fit to the $K^+\Sigma^0$ data only. 
It is obvious from Table \ref{tab:coupling_constant}
that the new data have a significant impact on the fit result. In general, including
the new data reduces the coupling strengths of exchanged particles.
This phenomenon can be understood from the fact that models which nicely fit
the $K^+\Sigma^0$ data tend to notoriously overpredict the $K^0\Sigma^+$ 
cross sections \cite{Mart:1995wu}. All coupling constants need to be readjusted
in order to reproduce the new $K^0\Sigma^+$ data.
Since both $g_{K\Lambda N}$ and $g_{K\Sigma N}$ coupling constants are fixed
to the SU(3) values, the $K^*$ coupling constants decrease by almost one
order of magnitude. The tensor coupling of $K_1$ becomes also smaller after the 
inclusion of new data. It has been known that the $K^*$ and $K_1$ exchanges 
strongly determine the shape of angular distribution of the differential
cross sections. It is important to note that if the Born terms dominate the
process then, due to the missing of $K^0$ intermediate state in the
$K^0\Sigma^+$ channel, the calculated differential cross sections show a
backward peaking behavior. However, as shown by the hadronic form factor
cut-off $\Lambda_{\rm Born}$ in Table \ref{tab:coupling_constant}, the model 
is not dominated by the Born terms. Therefore, only readjustment of the coupling constants is
required in order to fit the new $K^0\Sigma^+$ data. The $K^*$ coupling constants, 
which are responsible for the divergent behavior of cross sections at higher 
energies \cite{Mart:1999yc}, are strongly suppressed. Nevertheless, in view of
the large $\chi^2/N_{\rm dof}$ obtained after including the new data, further
improvement is inevitably required.

\begin{table}
\caption{Coupling constants ($g_i$) and hadronic form factor cut-offs ($\Lambda_i$)
  extracted from fits without (a) and with (b) the new $K^0\Sigma^+$ data.
  The Born coupling constants $g_{K\Lambda N}$ and $g_{K\Sigma N}$ are fixed
  to the SU(3) values.\label{tab:coupling_constant}}
\begin{center}
\begin{tabular}{lrr}
\hline\hline\noalign{\smallskip}
Coupling Constants & (a)~ & (b)~ \\\noalign{\smallskip}
\hline\noalign{\smallskip}
$g_{K\Lambda N}/\sqrt{4\pi}$ &  $-3.80$  &   $-3.80$\\
$g_{K\Sigma N}/\sqrt{4\pi}$  &   1.20  &    1.20\\
$g_{K^*K\gamma}~g^V_{K^*\Lambda N}/{4\pi}$ & 1.56 & $-0.22$\\
$g_{K^*K\gamma}~g^T_{K^*\Lambda N}/{4\pi}$  & 3.22 & 0.46\\
$g_{K_1K\gamma}~g^V_{K_1\Lambda N}/{4\pi}$  & $-5.00$ &  5.00\\
$g_{K_1K\gamma}~g^T_{K_1\Lambda N}/{4\pi}$  & $-1.86$ & $-0.57$\\
$g_{N^*(1650)N\gamma}~g_{K\Lambda N^*(1650)}/\sqrt{4\pi}$ & $-0.10$ & 0.06\\
$g_{N^*(1710)N\gamma}~g_{K\Lambda N^*(1710)}/\sqrt{4\pi}$ & $-0.25$ & 0.04\\
$g_{\Delta(1900)N\gamma}~g_{K\Sigma\Delta(1900)}/\sqrt{4\pi}$ & 0.11 & 0.04\\
$g_{\Delta(1910)N\gamma}~g_{K\Sigma\Delta(1910)}/\sqrt{4\pi}$ & 0.46 & 0.58\\
$g^{(1)}_{N^*(1720)N\gamma}~g_{K\Lambda N^*(1720)}/\sqrt{4\pi}$ & $-0.11$ & $-0.01$\\
$g^{(2)}_{N^*(1720)N\gamma}~g_{K\Lambda N^*(1720)}/\sqrt{4\pi}$ &  0.34 & $-0.71$\\
$g_{K_1^0K^0\gamma}~/~g_{K_1^+K^+\gamma}$ &  -~~ & $-0.12$\\
$\Lambda_{\rm Born}$ (GeV) & 0.53 & 0.55\\
$\Lambda_{\rm Res.}$ (GeV) & 1.38 & 1.07\\
\noalign{\smallskip}\hline\noalign{\smallskip}
$\chi^2/N_{\rm dof}$ & 2.36 & 4.14 \\
$N_{\rm data}$        & 676  & 818  \\
\noalign{\smallskip}\hline\hline
\end{tabular}
\end{center}
\end{table}

\begin{table}
\caption{Extracted masses ($M$) and widths ($\Gamma$) of resonances (in MeV) 
  from different models. Model 1 and Model 2 are described in the paper. 
  Both $K^+\Sigma^0$ and $K^0\Sigma^+$ data are used in this fit.
  \label{tab:float_masses}}
\begin{center}
\begin{tabular}{lcrrr}
\hline\hline\noalign{\smallskip}
Resonance & Mass  or & Original & Model 2 & Model 1 \\
          & Width   & value~ \cite{PDG} &  & \\\noalign{\smallskip}
\hline\noalign{\smallskip}
$S_{11}(1650)$ & $M$      & 1650 & 2167 & 1795 \\
               & $\Gamma$ & 150  & 186  & 158  \\
               & $M$      &   -~~&  -~~ & 2112 \\
               & $\Gamma$ &   -~~&  -~~ & 400  \\
$P_{11}(1710)$ & $M$      & 1710 & 1690 & 1680 \\
               & $\Gamma$ & 100  & 100  & 100 \\
$P_{13}(1720)$ & $M$      & 1720 & 2133 & 2141 \\
               & $\Gamma$ & 150  & 256  & 279 \\
$S_{31}(1900)$ & $M$      & 1900 & 1920 & 1900 \\
               & $\Gamma$ & 200  & 355  & 329 \\
$P_{31}(1910)$ & $M$      & 1910 & 1936 & 1800 \\
               & $\Gamma$ & 250  & 399  & 400 \\
\noalign{\smallskip}\hline\noalign{\smallskip}
$\chi^2/N_{\rm dof}$ & & 4.14 & 2.44 & 1.76 \\
\noalign{\smallskip}\hline\hline
\end{tabular}
\end{center}
\end{table}

Since there is no hint for a new resonance neither in the differential nor 
in the total cross section (see Figs. \ref{pic:modelsdiffwq} and \ref{pic:modelswqtot}), 
an arbitrarily inclusion of new resonances to improve the $\chi^2$ is by no means 
advocated. Instead, we left the masses and widths of
$N^*$ and $\Delta$ resonances as free parameters within certain ranges to be 
determined by the fit. The result is shown by Model 2 in Table \ref{tab:float_masses}, 
where we can see that the $\chi^2/N_{\rm dof}$ is significantly reduced to 2.44.
It is, however, especially interesting to see that both $S_{11}$ and $P_{13}$ masses are
shifted to higher values (2167 MeV and 2133 MeV, respectively), whereas those of other 
resonance states are relatively stable. For $P_{11}$ the fits prefer a width 
at the lowest value allowed. Reference \cite{Mart:2004ug} pointed
out that such a behavior could be an indication for the existence of another 
resonance with a relatively different mass. As a first step to check 
this conjecture, we put two $S_{11}$ resonances and leave their masses and widths to 
be determined by the fit. As shown by Model 1 in the same Table, we obtain 
from such a fit two $S_{11}$ signals with masses 1795 MeV and 2112 MeV, respectively, 
which seemingly supports the finding in Ref. \cite{Mart:2004ug}. 
This could be another indication that more ``missing resonances'' are 
required to explain kaon photoproduction process, a point which should 
be addressed in future studies.

\begin{figure}[h]
  \center\resizebox{0.4\textwidth}{!}{\includegraphics{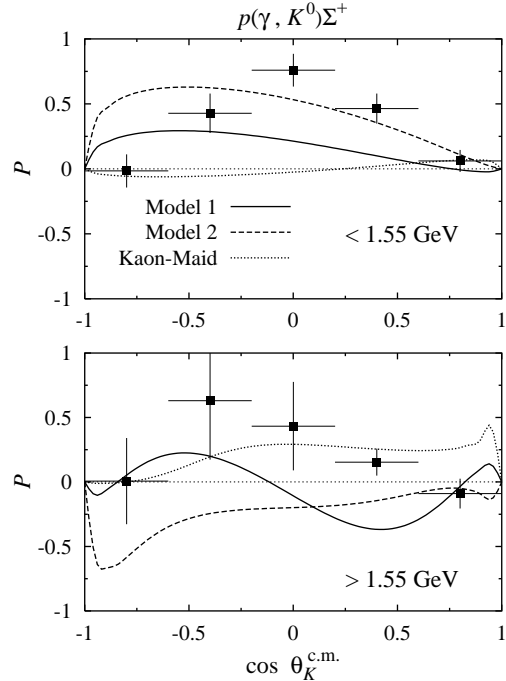}}
    \caption{As in Fig.\ \ref{pic:modelsdiffwq}, but for the $\Sigma^0$ recoil
      polarization.}
    \label{pic:modelspol}
\end{figure}

Figure \ref{pic:modelsdiffwq} compares the differential cross sections obtained
from different isobar models. It is evident from this figure that the KAON-MAID 
fit to the previous data is
unable to reproduce neither the shape nor the magnitude of differential cross
sections. In contrast to this, both Model 1 and Model 2 can fairly describe 
these new data up to some structures shown, except at low energies for backward 
angles. At forward directions Model 1 tends to produce more forward peaking cross 
sections at high energies than Model 2. The fact that Model 2 cannot reproduce
total cross section data at energies below 1900 MeV (Fig. \ref{pic:modelswqtot}) is 
due to the lack of resonances with $M\approx 1800$ MeV (see Table 
\ref{tab:float_masses}). In Model 1 the fitted mass of the first $S_{11}$ and that 
of the $P_{31}$ are in this region. We also note that in the case of the KAON-MAID 
fit on the previous data the divergent behavior of the total cross section at high 
energies is attributed to the large value of the $\Lambda_{\rm Born}$ cut-off (0.82 GeV). 
At this region a slight increment in total cross section is also observed in Model 1, 
but not in Model 2. Figure \ref{pic:modelspol} shows that neither model can reproduce the 
$\Sigma^+$ recoil polarization data which tend to peak at $\theta_K^{c.m.} \approx 90$ degrees. 
From the fit point of view this can be understood since the number of data is to scarce 
to compete with the cross sections one and the energy bins of the data are too broad 
to compare with a single energy prediction. On the other side, this result shows that
polarization data are still powerful to severely constrain the proliferating of the
models, provided that the corresponding accuracy can be significantly
improved.

\clearpage

\section{Summary}
\label{summary}
A new measurement of the reaction $\gamma p \rightarrow K^0 \Sigma^+$ 
carried out with the SAPHIR detector at ELSA is reported. 
The results comprise measurements of cross section and hyperon polarization as a 
function of kaon production angle and photon energy in the photon energy range from the 
reaction threshold up to $2.6$ GeV. The reaction cross section integrated over 
the angular range grows up to a photon energy around $\approx 1.4$ GeV and 
falls monotonously from there up to the highest measured energy. 
The reaction cross section is below that of $\gamma p \rightarrow K^+ \Sigma^0$ 
and varies less with photon energy and kaon production angle. The $\Sigma^+$ is 
polarized mainly in the angular region of $\cos\theta_K^{c.m.} \approx 0$. The data 
can be fairly well described within the framework of isobar model calculations.\\
{}\\
{}\\
We would like to thank the technical staff of the ELSA machine 
group for their invaluable contributions to the experiment. We 
gratefully acknowledge the support by the Deutsche Forschungsgemeinschaft 
in the framework of the Schwerpunktprogramm ``Investigation of the hadronic 
structure of nucleons and nuclei with electromagnetic probes'' 
(SPP 1034 KL 980/2-3).

\end{sloppypar}

% Non-BibTeX users please use

\end{document}